\documentclass[reprint,aps,pra,superscriptaddress,showpacs,nofootinbib]{revtex4-1}
\usepackage{blindtext}
\usepackage{graphicx}
\usepackage{dcolumn}
\usepackage{bm}
\usepackage{color}
\usepackage{ulem}
\usepackage{soul}
\usepackage{amsmath}
\usepackage[english]{babel}
\usepackage[symbol]{footmisc}

\begin{document}

\title{Performance of a Sagnac interferometer to observe vacuum optical nonlinearity}

\author{Aur\'elie Max Mailliet}
\email{mailliet@ijclab.in2p3.fr}
\author{Adrien E. Kraych}
\author{Fran\c{c}ois Couchot}
\author{Xavier Sarazin}
\author{Elsa Baynard}
\author{Julien~Demailly}
\author{Moana Pittman}
\affiliation{Université Paris-Saclay, CNRS/IN2P3, IJCLab, 91405 Orsay, France}
\author{Arache Djannati-Ata\"{i}}
\affiliation{Universit\'e Paris Diderot, CNRS/IN2P3, APC, 75013 Paris, France}
\author{Sophie Kazamias}
\affiliation{Université Paris-Saclay, CNRS/IN2P3, IJCLab, 91405 Orsay, France}
\author{Scott Robertson}
\affiliation{Institut Pprime, CNRS -- Université de Poitiers -- ISAE-ENSMA, 86073 Poitiers, France}
\author{Marcel Urban}
\affiliation{Université Paris-Saclay, CNRS/IN2P3, IJCLab, 91405 Orsay, France}


\begin{abstract}
In Quantum Electrodynamics, vacuum becomes a nonlinear optical medium: its optical index should be modified in the presence of intense external electromagnetic fields. The DeLLight project (Deflection of Light by Light) aims to observe this effect using intense focused femtosecond laser pulses delivered by LASERIX. The principle is to measure with a Sagnac interferometer the deflection of a low-intensity focused pulse (probe) crossing the vacuum index gradient induced by a high-intensity pulse (pump). 
A Sagnac interferometer working with femtosecond laser pulses has been developed for the DeLLight project. Compared to previous prototypes, the interferometer now includes the focusing of the probe beam in the interaction area.  
In this article, 
we measure and characterize the critical experimental parameters limiting the sensitivity of the interferometer, namely the extinction factor, the spatial resolution, and the waist at focus of the probe pulse. We discuss future improvements.
\end{abstract}

\maketitle



\section{Introduction}

The vacuum described by Quantum Electrodynamics (QED) should behave as a nonlinear optical medium, akin to all other optical media: the light velocity in vacuum should be reduced when the vacuum is stressed by intense electromagnetic fields. In other words, the vacuum optical index is expected to be increased when intense fields are applied to vacuum.  
This was predicted initially by Euler and Heisenberg~\cite{Euler-Heisenberg}, who derived an effective electromagnetic Lagrangian with nonlinear terms induced by the coupling of the fields with the electron-positron virtual pairs present in vacuum. This was later reformulated by Schwinger in the QED framework, showing that the Maxwell equations become nonlinear in the presence of intense constant (or slowly varying) fields~\cite{Schwinger}. 
Experimental efforts to observe a direct manifestation of a nonlinear optical effect in vacuum have mainly consisted in testing vacuum magnetic birefringence in the presence of an external  magnetic field, with an expected change of the vacuum optical index along the polarisation of the applied field~\cite{birefringence-PVLAS,birefringence-Toulouse,birefringence-Japan,birefringence-China}. However, no signal has yet been observed, and the best sensitivity to date has been achieved by the PVLAS experiment which is searching for vacuum birefringence with an external continuous magnetic field of 2.5~Tesla. After 100 days of collected data, the experiment reached a noise level ($1 \sigma$ confidence level) 7 times higher than the expected QED signal~\cite{birefringence-PVLAS}.

The DeLLight experiment (Deflection of Light by Light) seeks to measure optical nonlinearity in vacuum using the ultra high electric and magnetic fields contained in intense ultra short (femtosecondes) laser pulses. 
The current experimental developments are being carried out using the LASERIX facility (IJCLab, Université Paris-Saclay) which delivers femtosecond laser pulses of energy up to 2.5~J per pulse and a repetition rate of 10 Hz.
The initial idea was proposed in~\cite{DeLLight-2016}.
The experimental method and the calculation of the experimental sensitivity were described in~\cite{DeLLight-2021}.
The sensitivity depends upon three critical experimental parameters: the extinction factor of the interferometer, the spatial resolution of the measurement, and the probe beam size at focus inside the interferometer.
Results of a preliminary Sagnac interferometer prototype with unfocused femtosecond laser pulses were presented in~\cite{DeLLight-2021}. 

In this paper, we present the results of the sensitivity measurement obtained with the recent DeLLight pilot interferometer, including the focus of the probe beam and a reduction of the phase noise. 
We first recall in Section~\ref{sec:method-sensitivity} the experimental principle of the DeLLight measurement and summarize the expected sensitivity and the related critical parameters.
We then describe in Section~\ref{sec:setup} the pilot interferometer. 
The measurements of the extinction factor and the spatial resolution are presented in Sections~\ref{sec:extinction-factor} and~\ref{sec:spatial-resolution}, respectively. 
The challenge to reach a small waist at focus and the resulting optimisation of the parameters are discussed in Section~\ref{sec:minimum-waist-at-focus}. 
We then conclude in Section~\ref{sec:conclusion}.

\section{Experimental method}
\label{sec:method-sensitivity}

\subsection{Interferometric measurement}

\begin{figure}[!h]
    \centering
    \includegraphics[width=1\columnwidth]{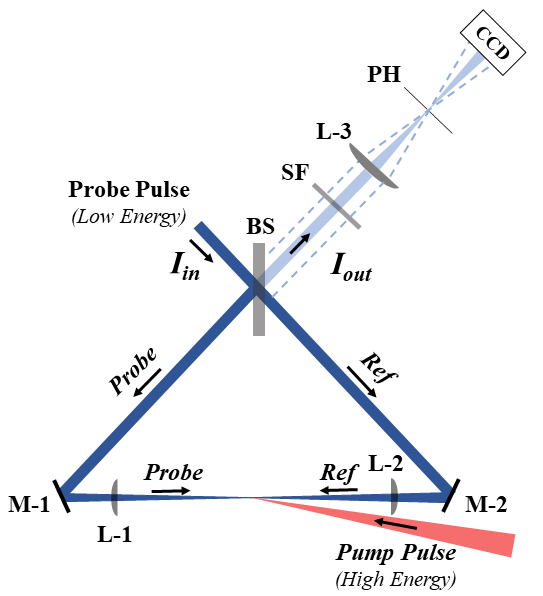}
    \caption{Optical scheme of the DeLLight Sagnac interferometer. The pump beam is in red. The probe and reference pulses counter-propagating inside the interferometer are in dark blue. The destructive interference is in light blue. The back-reflections on the rear side of the beamsplitter, used for the beam pointing suppression, are presented in dashed line. See text for details.}
    \label{fig:schema_pilot_experiment}
\end{figure}

The DeLLight interferometric method to measure the deflection of a probe laser pulse by an intense pump laser pulse is as follows, and schematized in Figure~\ref{fig:schema_pilot_experiment}.
A low-intensity laser pulse, of intensity $I_{in}$ and duration $\Delta t$, is sent into a Sagnac interferometer via a 50/50 beamsplitter {\it BS}, generating two daughter pulses (referred to as probe and reference pulses) that circulate in opposite directions around the interferometer. Both pulses are focused in the interaction area via two optical lenses, {\it L-1} and {\it L-2}, inserted in the interferometer, with a minimum waist at focus $w_0$. 
The two counter-propagating pulses are in opposite phase in the dark output of the interferometer and interfere destructively.  
The beamsplitter is never perfectly symmetric in reflection and transmission, and the extinction of the interferometer is never total. 
Therefore, a residual interference signal in the dark output, of intensity $I_{out}$, is measured by a CCD camera.
The degree of extinction of the interferometer is characterized by the extinction factor $\mathcal{F}$, defined as $\mathcal{F} = I_{out}/I_{in}$. 
The stronger the extinction is, the lower the value of $\mathcal{F}$ is.

The intense pump pulse of energy $E_{pump}$, and same duration $\Delta t$ as the probe pulse,
is now focused in the interaction area with a minimum waist at focus $W_0$, in time coincidence and counter-propagating with respect to the probe pulse. Its nonlinear coupling with the probe generates in the vacuum an optical index gradient $\delta n_{\mathrm{QED}}$ proportional to the intensity profile of the pump. 
As calculated in~\cite{Robertson-2019}, the index variation $\delta n_{\mathrm{QED}}$ is maximal when the pump and the probe are counter-propagating and becomes null when they are co-propagating. It is also shown in~\cite{Robertson-2019} that $\delta n_{\mathrm{QED}}$ is maximal when the polarisations of the pump and the probe are perpendicular, and is $\frac{4}{7}$ smaller when the polarisations are parallel. 

The probe is then refracted by the index gradient $\delta n_{\mathrm{QED}}$, while the reference pulse is not in time coincidence with the pump and is therefore unaffected. The axis of the pump is vertically shifted with respect to the axis of the probe, by a distance $b$ named the impact parameter, so that the perturbation of the probe is asymmetric and its mean deflection angle $\delta \theta_{\mathrm{QED}}$ is nonzero. 
For Gaussian pulses, the deflection is maximal when the impact parameter $b$ is equal to $b=b_{\mathrm{opt}}=\sqrt{W_0^2+w_0^2}/2$.
After re-collimation by the second lens, the refracted probe pulse is vertically shifted with respect to the unrefracted reference pulse by an average distance $\delta y_{\mathrm{QED}} = f \times \delta \theta_{\mathrm{QED}} $, where $f$ is the focal length. The probe is also delayed by the index gradient induced by the pump, generating a phase delay signal $\delta \psi_{\mathrm{QED}}$. 

When the deflected probe pulse interferes destructively with the unperturbed reference pulse, the transverse intensity profile of the interference signal in the dark output is vertically shifted by a distance $\Delta y_{\mathrm{QED}}$, which is amplified as compared to the would-be direct signal $\delta y_{\mathrm{QED}}$ obtained when using a standard pointing method. 
We will show in the following that the amplification factor,
defined as $\mathcal{A} = \Delta y_{\mathrm{QED}}/\delta y_{\mathrm{QED}}$, scales as $\mathcal{F}^{-1/2}$. 
In other words, the greater the extinction is, the greater the amplified signal is.
This amplification of the deflection signal is the basis for the use of the Sagnac interferometer.

\subsection{Amplification factor of the deflection signal}

Let us take a closer look at how the amplification factor $\mathcal{A}$ depends upon the asymmetry of the beamsplitter and the quality of the interferometer.

The intensity asymmetry of the beamsplitter is parameterized by the small quantity $\delta a$, as
\begin{eqnarray}
& T  =  \frac{1}{2} \left(1+\delta a\right) \,, \qquad R  =  \frac{1}{2} \left(1-\delta a\right)  \,,
\label{eq:imperfect-beamsplitter}
\end{eqnarray}
where  $T$ and $R$ are the transmission and reflection coefficients in intensity of the beamsplitter. 
Morever, the two counter-propagating pulses interfering in the dark output are never perfectly in opposite phase, either because of a possible residual constant phase induced by the beamsplitter itself, denoted $\delta \phi_0$, or because of a surface phase noise, denoted $\delta \phi_s(x,y)$, induced by the surface defects of the optics inside the interferometer and which depends upon the transverse coordinates $(x, y)$.
We assume that the transversal dependence of $\delta a$ and $\delta \phi_0$ is negligible and both parameters can be considered as uniform.
We denote $\delta \phi(x,y) = \delta \phi_0 + \delta \phi_s(x,y)$ the total phase noise in the dark output.

In the absence of the pump (referred to as an {\it OFF} measurement), the transverse intensity profile $I_{OFF}(x,y)$ of the interference of the probe and reference pulses measured in the dark output, is given by~\cite{DeLLight-in-air}:
\begin{equation}
    I_{OFF}(x,y) = \left( (\delta a)^2 + \left( \delta \phi(x,y) \right)^2 \right) \times I_{in}(x,y) 
    \label{eq:Ioff}
\end{equation}
and the extinction factor $\mathcal{F}$ is thus equal to:
\begin{equation}
\mathcal{F} = (\delta a)^2 + \left(\delta \phi(x,y) \right)^2
\label{eq:extinction-factor-complet}
\end{equation}

When the pump is now active and interacts with the probe (referred to as an {\it ON} measurement), the transverse intensity profile $I_{ON}(x,y)$ of the interference is equal to~\cite{DeLLight-in-air}:
\begin{align}
\begin{split}\label{eq:Ion-general}
    I&_{ON}(y) =  (\delta a)^2 \times I_{in} \left(y - \frac{(\delta a-1)}{2\delta a} \times\delta y_{\mathrm{QED}} \right)\\
                    & + \left(\delta \phi(y) + \frac{\delta \psi_{\mathrm{QED}}}{2} \right)^2 \times \left(1-(\delta a)^2\right) \times I_{in}\left(y-\frac{\delta y_{\mathrm{QED}}}{2}\right)
\end{split}
\end{align}
For a high extinction, $\delta a \ll 1$ and Equation~\ref{eq:Ion-general} becomes
\begin{align}
\begin{split}\label{eq:Ion}
    I_{ON}(x, y) ={}&  (\delta a)^2 I_{in} \left(x,y + \frac{\delta y_{\mathrm{QED}}}{2\delta a} \right)\\
                    & + \left(\delta \phi(x,y) + \frac{\delta \psi_{\mathrm{QED}}}{2} \right)^2  I_{in}\left(x,y-\frac{\delta y_{\mathrm{QED}}}{2}\right)
\end{split}
\end{align}

In vacuum, the phase signal $\delta \psi_{\mathrm{QED}}$ is extremely small and $\delta \psi_{\mathrm{QED}} << \delta \phi$. Morever, when the contribution of the phase noise is at first order negligible compared to the asymmetry parameter $(\delta \phi)^2  \ll (\delta a)^2$, then Equations~(\ref{eq:Ioff}), (\ref{eq:extinction-factor-complet}) and (\ref{eq:Ion}) become
\begin{eqnarray}
    I_{OFF}(x,y) & = & (\delta a)^2 \times I_{in}(x,y) \\
    I_{ON}(x,y) & = & (\delta a)^2 \times I_{in}\left(x,y + \frac{\delta y_{\mathrm{QED}}}{2 \delta a}  \right) \\
    \mathcal{F} & = & (\delta a)^2
    \label{eq:IoffIon-simplified}
\end{eqnarray} 
As a result, when the probe is deflected by its interaction with the pump, the barycenter of the interference intensity profile in the dark output is then vertically shifted by a distance $\Delta y_{\mathrm{QED}} = \delta y_{\mathrm{QED}} /(2\delta a)$. 
The vertical displacement signal $\Delta y_{\mathrm{QED}}$ is thus amplified as compared to the would-be direct signal $\delta y_{\mathrm{QED}}$  when using a standard direct pointing method. The amplification factor defined as $\mathcal{A}=\Delta y/\delta y$ is then inversely proportional to $\sqrt{\mathcal{F}}$: 
\begin{eqnarray}
\mathcal{A} =  - \frac{1}{2\delta a} = - \frac{1}{2\sqrt{\mathcal{F}}}
\end{eqnarray}
For an extinction factor $\mathcal{F}= 4 \times 10^{-6}$, as measured in our current pilot interferometer (see section~\ref{sec:extinction-factor}), the displacement signal is amplified by a factor $|\mathcal{A}| = 250$.

\subsection{Expected sensitivity}

\begin{figure}[!h]
    \centering
    \includegraphics[width=1\columnwidth]{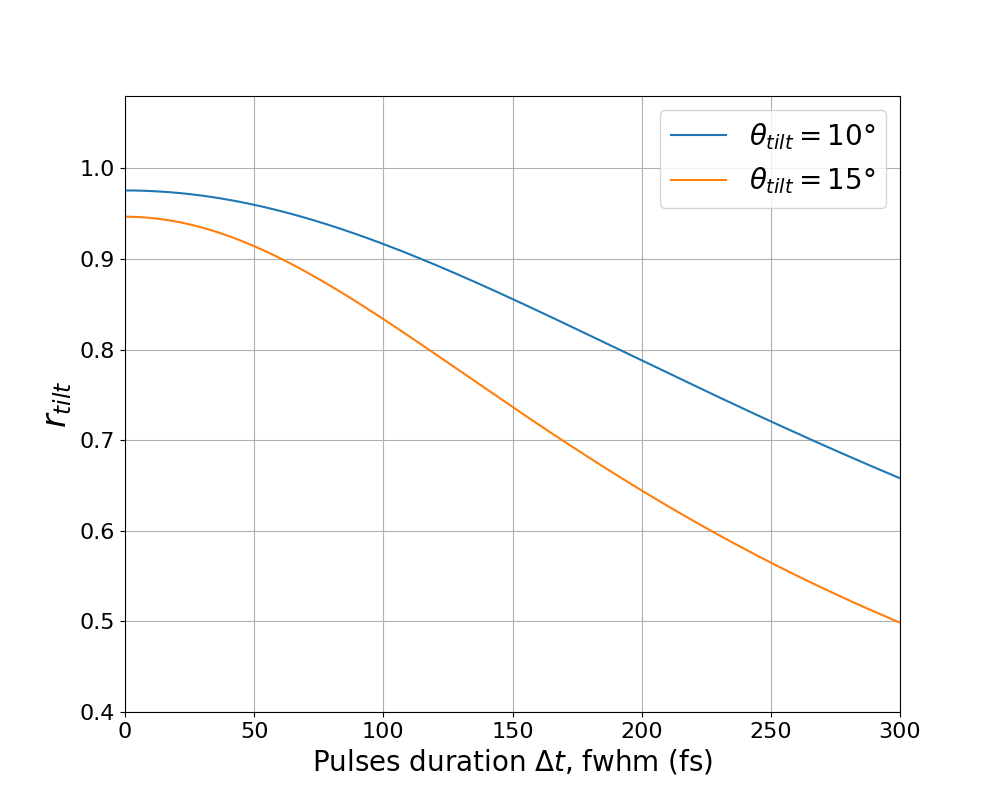}
    \caption{Correction factor $r_{\mathrm{tilt}}$ as a function of the duration $\Delta t$ (fwhm) of the pump and probe pulses, for two different values of the tilt angle between the probe and the pump beam. 
    The expected signal is around 20$\%$ smaller with a pulse duration of 200~fs, compared with 30~fs.}
    \label{fig:rtilt}
\end{figure}

As calculated in~\cite{DeLLight-2021}, when $b=b_{\mathrm{opt}}$ and $(\delta \phi)^2  \ll (\delta a)^2$, the expected QED signal $\Delta y_{\mathrm{QED}}$ (in pm) is equal to:  
\begin{equation}
    \Delta y_{\mathrm{QED}}(\mathrm{pm}) = 9.0 \times  \frac{E_{pump}(\mathrm{J}) \times f(\mathrm{m})}{(w_0^2(\mathrm{\mu m})+W_0^2(\mathrm{\mu m}))^{3/2} \times \sqrt{\mathcal{F}} } \times r_{\mathrm{tilt}}
    \label{eq:expected-signal}
\end{equation}
where $r_{\mathrm{tilt}}$ is a correction factor which takes into account the tilt angle $\theta_{\mathrm{tilt}}$ between the probe and the pump beam in the horizontal plane. Its value, calculated in~\cite{DeLLight-2021}, depends on the time duration $\Delta t$ (fwhm) of the pump and probe pulses and is shown in Figure~\ref{fig:rtilt}, for a tilt angle $\theta_{\mathrm{tilt}}=10^{\circ}$ corresponding to the realistic minimum angle taking into account mechanical constraints. The dependence of the tilt angle on the expected signal is presented in~\cite{DeLLight-2021}.

With an energy $E_{pump}=2.5$~J for the pump pulse (as designed for the LASERIX laser), a waist at focus $W_0=w_0=5 \ \mu$m, and as 
justified later in this article, assuming an optimal extinction factor $\mathcal{F}= 4 \times 10^{-6}$ (an amplification factor $\mathcal{A} = 250$), a focal length $f=250$~mm, a tilt angle $\theta_{\mathrm{tilt}}=10^{\circ}$ and an optimal pulse duration $\Delta t = 180$~fs for both the pump and the probe pulses,
the expected deflection signal is then $\Delta y_{\mathrm{QED}} = 6$~pm. It corresponds to a deflection angle $\delta \theta_{\mathrm{QED}} = 0.13$~prad.
The signal is measured by alternating measurements with positive $+b_{\mathrm{opt}}$ and negative $-b_{\mathrm{opt}}$ impact parameters of the pump beam (Up-Down measurements), corresponding to a measured signal $\Delta y(\mathrm{up-down}) = 2 \times \Delta y_{\mathrm{QED}} = 12$~pm. 
With a spatial resolution of the CCD readout of $\sigma_y = 15$~nm (corresponding to the shot noise of commercial CCD cameras, as measured in section~\ref{sec:spatial-resolution}), and with the 10~Hz repetition rate of LASERIX (corresponding to 5~Hz Up-Down measurement frequency), the expected sensitivity (1$\sigma$ confidence level) to measure the QED signal could be reached after about 4~days of collected data. 

So, regardless the features of the laser that delivers the pump pulses, the capacity to observe a signal depends upon three critical parameters of the interferometer: the extinction factor $\mathcal{F}$, the spatial resolution $\sigma_y$, and the pump and probe beam sizes at focus $w_0$, with the following goal values: $\mathcal{F}= 4 \times 10^{-6}$, $\sigma_y = 15$~nm and $w_0=5 \ \mu$m.

\section{Description of the DeLLight pilot interferometer}
\label{sec:setup}

A Sagnac interferometer, named pilot interferometer, was developed in order to measure and characterize the three critical experimental parameters: the extinction factor of the interferometer,
the spatial resolution of the interference signal measurement, and the probe beam size at focus inside the interferometer.
A simplified scheme of the DeLLight pilot interferometer setup is given in Figure~\ref{fig:schema_pilot_experiment}.
The Sagnac interferometer is in a right-angled isosceles triangle configuration, composed of a beamsplitter {\it BS}, two mirrors {\it M-1} and {\it M-2}, and two lenses {\it L-1} and {\it L-2} to focus the pulses in the interaction area. 

Before entering the interferometer, the transverse waist of the incident beam is set to $w \simeq 1$~mm by an afocal telescope composed of two lenses and a pinhole at focus, which acts as a spatial filter in order to obtain a nearly Gaussian transverse intensity profile.
For the measurements presented in this article, the duration of the laser pulses is about 70~fs. The central wavelength is about 815~nm with a bandwidth of about 40~nm.
The energy of the incident beam may vary from to 2 to 20~$\mu$J per pulse, with a repetition rate of 10~Hz.
The $p$ polarisation of the incident beam is precisely adjusted just before entering the Sagnac interferometer, by using a motorised half-wave plate rotation stage. 
This polarisation adjustment is done in order to maximize the extinction in the dark output of the interferometer. 

The beam is then split in two by the 50/50 femtosecond p-pol beamsplitter BS (Semrock FS01-BSTiS-5050P-25.5): the transmitted pulse corresponds to the reference (Ref) and the reflected pulse corresponds to the probe (Probe). The interferometer is formed by the beamsplitter (BS), two dielectric mirrors positioned at a 22.5° incident angle (M-1 and M-2), and two best form spherical lenses (L-1 and L-2) of focal length 100 mm (Thorlabs LBF254-100-B), which focus both Probe and Ref in the interaction area. 
The transverse intensity profile measured at focus is Gaussian with a minimum waist $w_0 \simeq 25 \ \mu$m.
The beamsplitter BS, the mirror M-2 and the lens L-2 of the interferometer are controlled by static piezoelectric adjusters with nanometric accuracy. 
The thickness of the beamsplitter is 3~mm.
An anti-reflective coating has been deposited on the rear side of the beamsplitter.
The high reflective coating ($R = 99.99 \, \%$) of M-1 and M-2 was produced by the Laboratoire des Matériaux Avancés (LMA, IP2I-Lyon).

The interference signal of the Probe and Ref is read in the dark output of the interferometer by a CCD camera (Basler acA3088-16gm, pixel size 5.84 × 5.84 $\mu{\rm m}^{2}$), which is installed outside the vacuum chamber.  
A spatial filter is placed in the dark output in front of the CCD in order to suppress the interference phase noise induced by diffusion on the surface defects of the optics inside the interferometer. 
The spatial filter consists of a focusing best form optical lens L-3 of focal length $f=200$~mm and a pinhole {\it PH} of diameter 200~$\mu$m at focus. 
The CCD camera is placed at exactly $2f$ so as to have a magnification of 1. 
An interferential multi-layer dielectric spectral filter {\it SF} of spectral width $\Delta \lambda = 3$~nm, centred at 808~nm, is  optionally installed in the dark output before the spatial filter. By rotating the incident angle of the spectral filter, we can select the wavelength from 808 down to 800~nm to optimise the extinction factor in the dark output.

\section{Extinction in the dark output of the interferometer}
\label{sec:extinction-factor}

\subsection{Measurement of the extinction factor}

\begin{figure*}
    \includegraphics[scale=0.35]{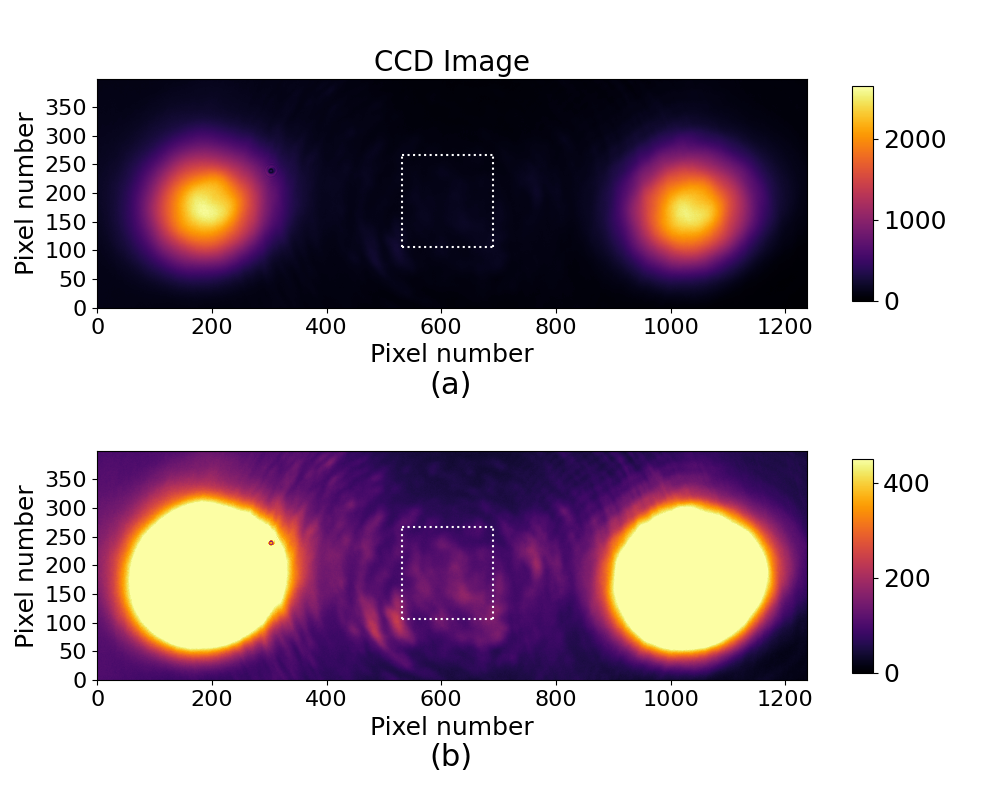}
    \includegraphics[scale=0.35]{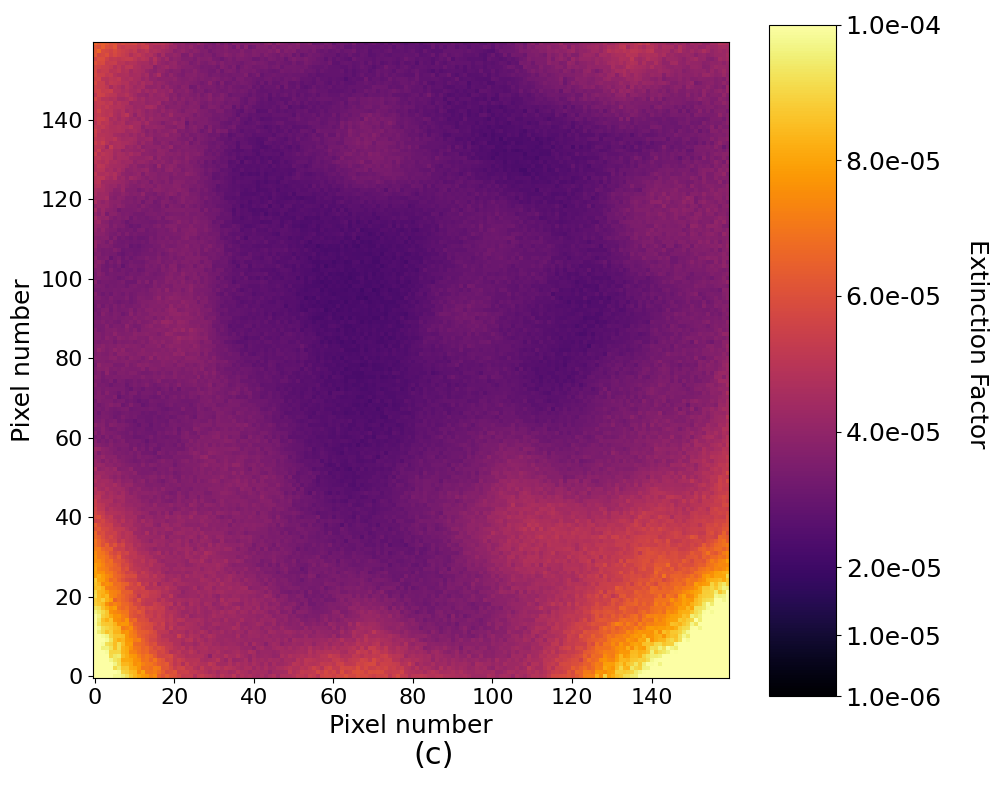}
    \caption{(a) and (b) CCD images recorded in the dark output of the DeLLight interferometer at maximal extinction, with a spectral filter and without spatial filter (the white dotted area corresponds to the Region of Interest (RoI) of the expected interference signal): (a) raw intensity profile; (b) same image with saturated back-reflections in order to  visualise the residual intensity pattern of the interference signal. (c) Distribution of the extinction factor in the RoI shown in the left panel.}
    \label{fig:extinction-1}
\end{figure*}

A typical transverse intensity profile of the interference signal, recorded by the CCD camera in the dark output of the interferometer after optimization of the extinction factor, is presented in Figure~\ref{fig:extinction-1}. The incident energy of the laser pulses entering the interferometer is about 20~$\mu$J. This image has been recorded with the spectral filter placed in the dark output of the interferometer. However, no spatial filter is installed here. 
Two spots, with approximately equal intensities $I_{\rm AR}$, are clearly observed on opposite lateral sides of the central destructive interference signal. 
As detailed in Appendix~B of reference~\cite{DeLLight-in-air}, and as illustrated in dashed lines in Figure~\ref{eq:extinction-factor-complet}, one spot corresponds to the direct image of the incident intensity after a single reflection on the rear side of the beamsplitter, while the second spot is a superposition of four distinct reflected waves.
The intensity of the direct reflection is $I_{\rm AR} = R_{\mathrm{AR}}/2 \times I_{in}$, where $R_{\mathrm{AR}}$ is the reflectivity coefficient (in intensity) of the anti-reflective coating deposited on the rear side of the beamsplitter.
It has been measured with a photodiode and a set of calibrated neutral densities, and its value is $R_{\mathrm{AR}} = (1.1 \pm 0.1) \times 10^{-3} $. 

\begin{figure*}
    \includegraphics[scale=0.35]{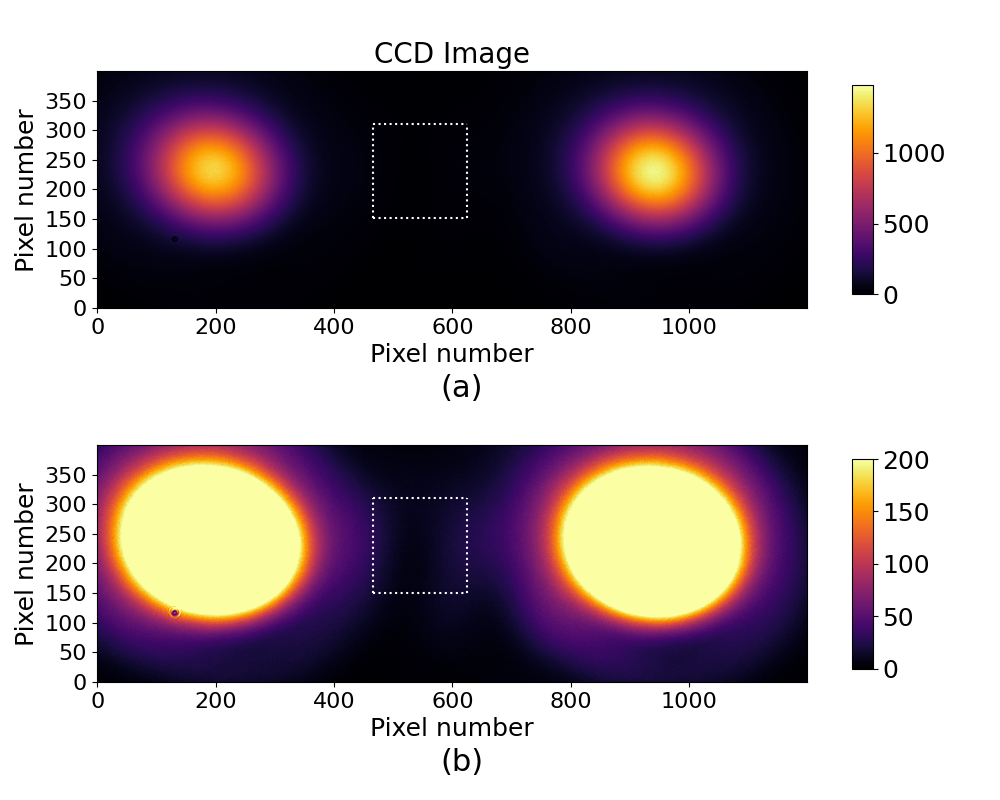}
    \includegraphics[scale=0.35]{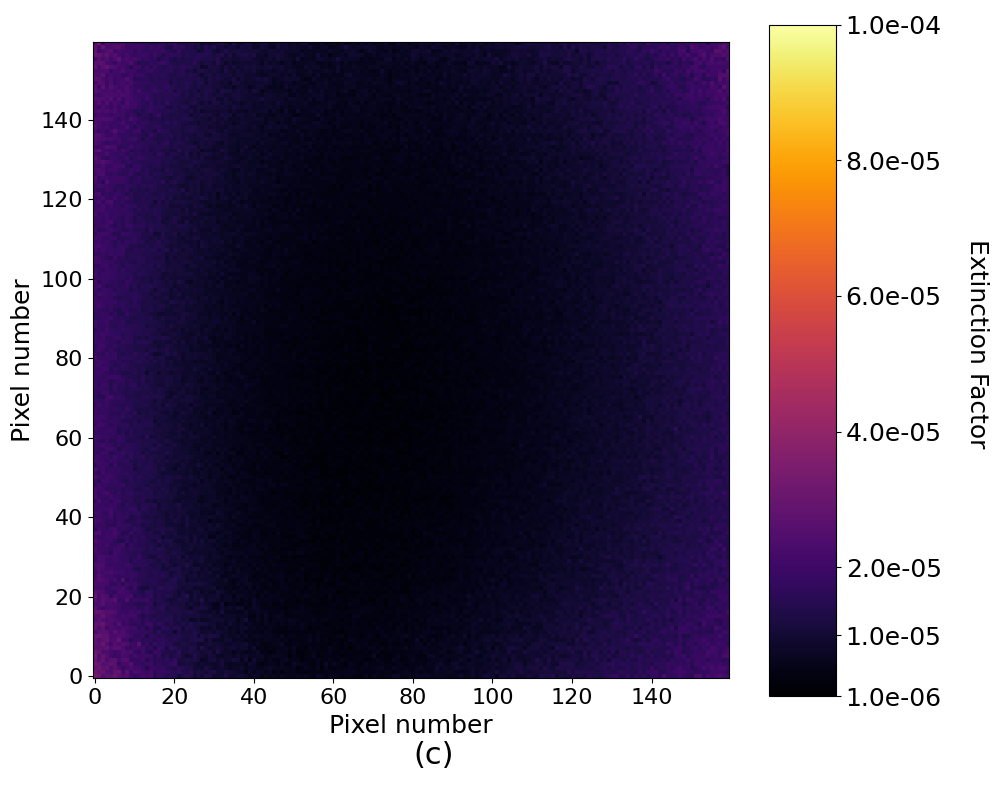}
    \caption{Same as Figure~\ref{fig:extinction-1} but with both a spatial filter and a spectral filter in the dark output of the interferometer.}
    \label{fig:extinction-2}
\end{figure*}

The interference signal is located in the central part, delimited by the dotted white square in Figure~\ref{fig:extinction-1}. 
The residual phase noise of the interference signal is shown in the lower left image of Figure~\ref{fig:extinction-1} by increasing the sensitivity scale of the display. 
Two types of phase noise pattern can be distinguished for the interference signal: interference rings with large transverse size and hot spots in the central area of the expected intensity signal. 
The diffraction rings correspond to a surface defect (flatness) on the mirrors at large scale (low spatial frequency), compatible with the surface flatness specifications of $\lambda/10$. 
The hot spots in the central area correspond to small sized surface defects due to roughness (high spatial frequency).

The extinction factor $\mathcal{F}$ of the interferometer is then calculated by measuring the intensity of the
interference signal $I_{out}$ relative to the intensity $I_{\mathrm{AR},1}$ of the direct back-reflection on the rear side of the beamsplitter:
\begin{equation}
    \mathcal{F} = \frac{I_{out}}{I_{\mathrm{AR}}} \times \frac{R_{\mathrm{AR}}}{2}
\end{equation}
The extinction factor is plotted in the right panel of Figure~\ref{fig:extinction-1} for each pixel $(i,j)$ in the delimited square region of interest (represented by the dashed white square), where one can observe the hot spots of the phase noise pattern $\delta\phi(x,y)$. 
The extinction factor measured in the central area is about $ \mathcal{F} \simeq 4 \times 10^{-5}$.
From Eq.~(\ref{eq:extinction-factor-complet}), it corresponds to a difference of phase  which is at maximal  $\delta \phi \le 6$~mrad (i.e when $\delta a \ll \delta \phi$), and is equivalent to a difference in the optical path lengths lower than $\delta l \le \delta \phi \times \lambda_0 /(2\pi) \simeq 8 \ \mathring{\mathrm{A}}$. Such value is within the tolerance of the surface quality of the lenses and the beamsplitter used here.


The extinction is then measured by adding the spatial filter in the dark output in front of the CCD camera. 
The recorded image of the extinction with the spatial filter is presented in Figure~\ref{fig:extinction-2}. 
The phase noise pattern induced by the diffusion is now well suppressed by the spatial filter. The extinction factor measured in the central area is now $ \mathcal{F} = 3 \times 10^{-6}$. 
The extinction is limited by the beamsplitter asymmetry $\delta a = 1.7 \times 10^{-3}$.

\begin{figure*}
    \includegraphics[scale=0.35]
    {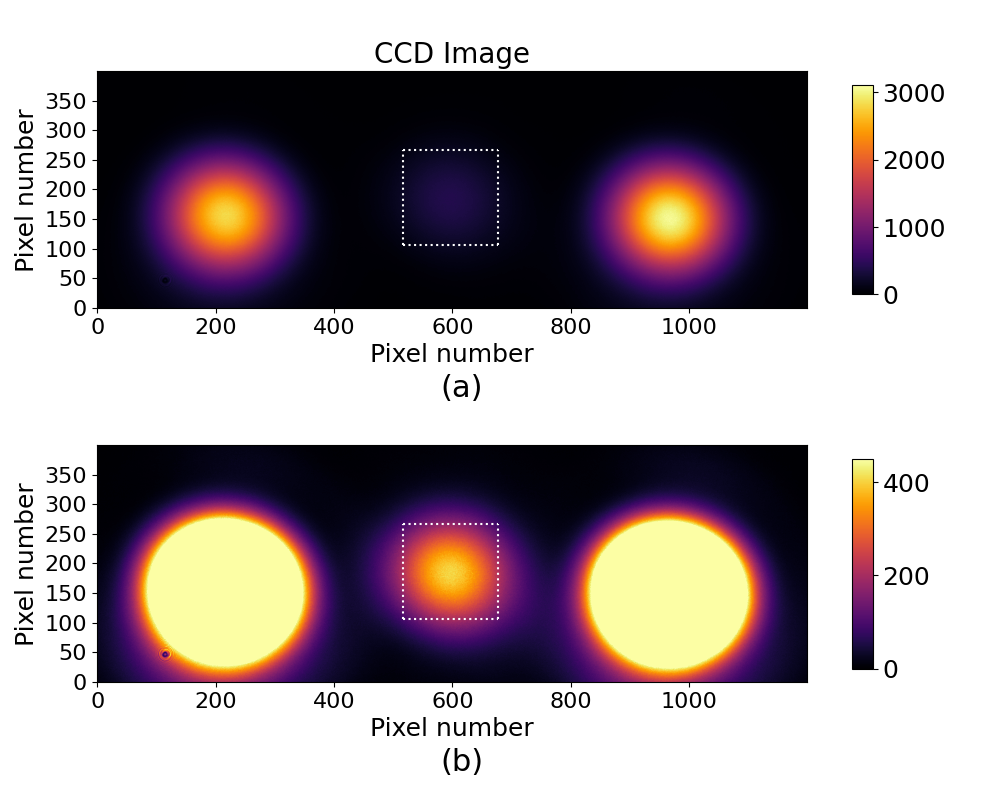}
    \includegraphics[scale=0.35]
    {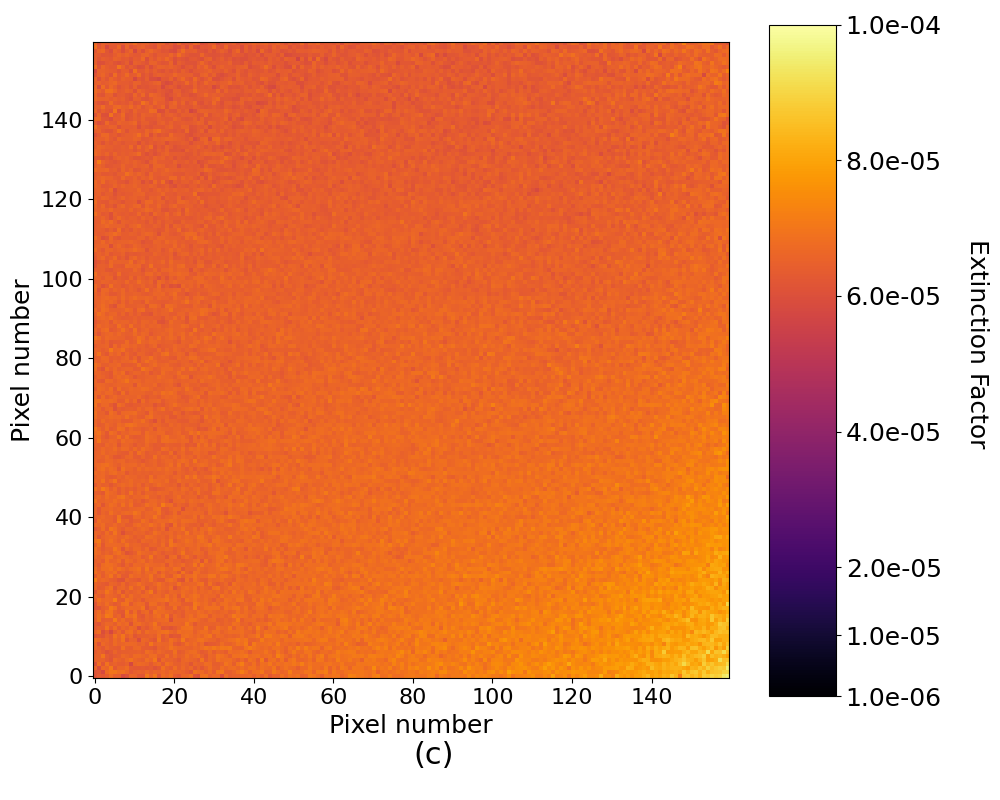}
    \caption{Same as Figure~\ref{fig:extinction-1} with a spatial filter and no spectral filter in the dark output of the interferometer.}
    \label{fig:extinction-3}
\end{figure*}


In order to measure the influence of the broadband spectrum of the laser, the extinction is then measured without spectral filter, i.e. with the full wavelength spectrum of the laser pulse. The spectral filter is replaced by a set of neutral densities which are placed in front of the CCD camera, in order to match the intensity attenuation provided by the spectral filter. The spatial filter is kept installed in front of the CCD camera.  
The result is presented in Figure~\ref{fig:extinction-3}. The interference signal is now clearly observed, with a limited extinction factor $ \mathcal{F} \simeq 6 \times 10^{-5}$ which corresponds to an average beamsplitter asymetry $\delta a = 8 \times 10^{-3}$.

\subsection{Development of new beamsplitters}

The goal value of the extinction factor $ \mathcal{F} = 4 \times 10^{-6}$ has been successfully achieved thanks to the insertion of a spatial filter. However it has been measured with only a part of the spectrum. A new beamsplitter with a dielectric coating ensuring the goal extinction factor over the full spectrum is therefore required.
Moreover, high photon statistics in the interference signal in the dark output is needed to reach a good spatial resolution (see Section~\ref{sec:spatial-resolution}) while maintaining the high extinction factor. 
To do this, the intensity of the incident pulses entering the interferometer must be increased.
However, the intensity of the back-reflections would then saturate the CCD camera. Additionally, the tail of the back-reflection intensity distributions would pollute the interference signal. 

The development of a new beamsplitter with a larger thickness (in order to increase the distance between the interference signal and the back reflections), and which includes a dedicated anti-reflective coating with lower back reflectivity coefficient,  is underway at the Laboratoire des Matériaux Avancés (LMA, IP2I Lyon, IN2P3). A first beamsplitter prototype (6.35~mm thickness) has been produced using an Ion Beam Sputtering (IBS) technique, and we have measured over the full spectrum an excellent reflectivity coefficient of the anti-reflective coating $R_{AR} = 10^{-5}$, but a limited extinction factor $\mathcal{F} = 10^{-4}$. A beamsplitter with a lower extinction factor is in development.

\subsection{Limitations of the extinction factor}
\label{sec:limitaion-extinction-factor}

The optical Kerr effect inside the fused silica substrate of the beamsplitter may limit the extinction factor of the interferometer. 
Indeed, the intensities of the probe (reflected) pulse $I_r$ and the reference (transmitted) pulse $I_t$ are different when they first pass through the beamsplitter, due to the beamsplitter asymmetry $\delta a$ and also to the fact that the probe pulse first reflects on the beamsplitter and the two mirrors and passes through the two lenses before passing through the beamsplitter\footnote{The Kerr effect is negligible in the second transmission of the beamsplitter since the two pulses interfere destructively. Therefore the intensity is strongly reduced.}. Noting $R_m=1-T_m$ the reflectivity coefficient of the mirrors and $R_{AR,L}$ the reflectivity coefficient of the anti-reflective coating of the lenses and $I_{\mathrm{in}}$ the intensity of the incident pulse, we have:
\begin{align}
\begin{split}\label{eq:delta_a}
I_t & =  \frac{I_{\mathrm{in}}}{2} \times (1+\delta a) \\
I_r & =  \frac{I_{\mathrm{in}}}{2} \times (R_m)^2 \times (1-R_{AR,L})^2 \times (1-\delta a) \\
& \simeq  \frac{I_{\mathrm{in}}}{2} \times (1-2T_m) \times (1-2 R_{AR,L}) \times (1-\delta a) 
\end{split}
\end{align}
The difference of intensity $\delta I$ between the probe and reference pulses when passing for the first time through the beamsplitter is then:
\begin{equation}
    \delta I = I_t - I_r \simeq I_{\mathrm{in}} \times (\delta a + T_m  + R_{AR,L})
    \label{eq:NL_dI}
\end{equation}
In the pilot interferometer, $T_m \simeq 10^{-4} \ll \delta a$ and with an appropriate anti-reflective coating on the lenses, we can also neglect $R_{AR,L} \ll \delta a$. Equation~(\ref{eq:NL_dI}) becomes simply $\delta I \simeq I_{\mathrm{in}} \times \delta a$.
This difference of intensity $\delta I$ generates a relative difference of the optical index in the beamsplitter $\delta n$ induced by the optical Kerr effect in the silica substrate, such as 
\begin{equation}
    \delta n \simeq I_{\mathrm{in}} \times \delta a \times n_2
    \label{eq:delta_n}
\end{equation}
where $n_2\simeq 3 \times 10^{-16}$~cm$^2$/W is the Kerr index in fused silica. 
This produces a phase shift $\delta \Phi_K$ between the probe and the reference pulses in the dark output, equal to:
\begin{equation}
\delta \Phi_K = \frac{2 \pi z}{\lambda} \times \delta n \simeq \frac{2 \pi e}{\lambda} \times n_2 \times I_{\mathrm{in}} \times \delta a
   \label{eq:phase-shift}
\end{equation}
where $\lambda$ is the wavelength of the laser, $z$ is the travelled distance by the beam in the beamsplitter and  $e$ the thickness of the beamsplitter ($z = 1.15 \times e \simeq e$ for an incident angle of 45°). 
This extra phase limits the extinction factor given now by 
\begin{eqnarray}
    \mathcal{F}_K & = & (\delta a)^2+(\delta \Phi_K)^2 \nonumber \\ 
    & = & \mathcal{F} \times \left(1+\left(\frac{2 \pi e \times n_2}{\lambda}\right)^2  \times I_{\mathrm{in}}^2 \right) 
    \label{eq:extinction-vs-intensity}
\end{eqnarray}
where $\mathcal{F} = (\delta a)^2$ is the extinction factor only limited by the beamsplitter asymmetry $\delta a$. 

\begin{figure}[h!]
    \centering
    \includegraphics[scale=0.35]{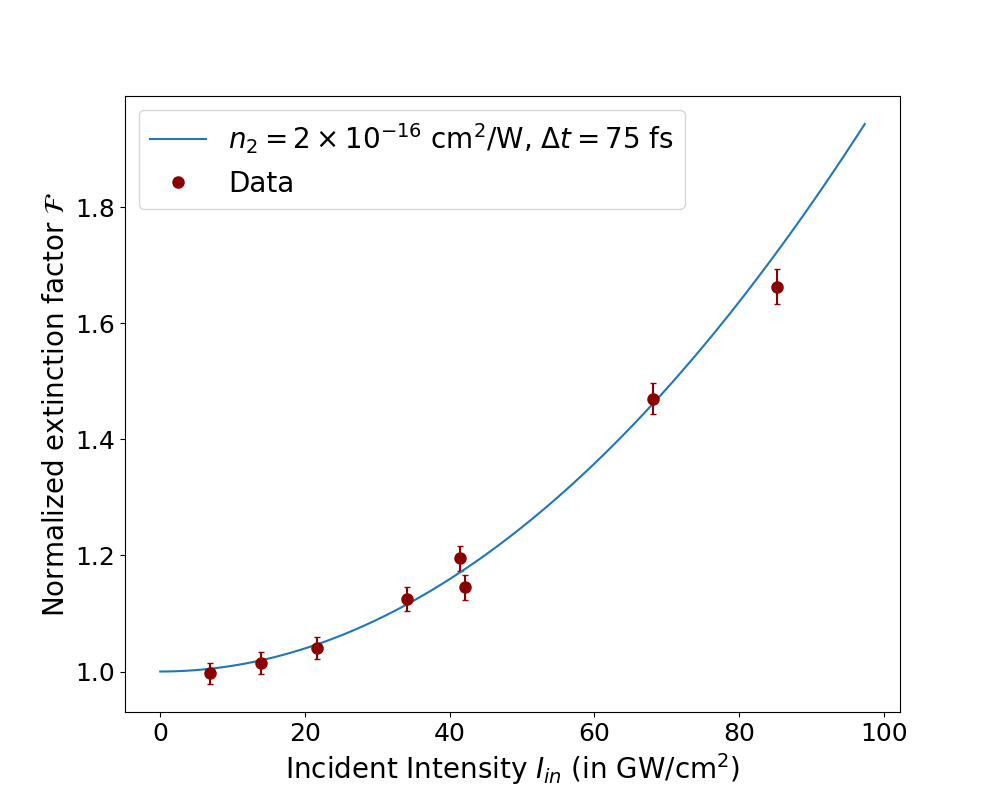}
    \caption{Extinction factor $\mathcal{F}$ as a function of the intensity $I_{\mathrm{in}}$ of the incident pulse entering the interferometer. The extinction factor has been normalised to 1 for the lowest intensity value. Measurements have been performed with a beamsplitter of thickness $e=6.35$~mm, a incident beam size $w=1$~mm and a pulse duration $\Delta t \simeq 70$~fs. Data (in red dot) are in good agreement with the expected value (in blue curve) calculated using Equation~(\ref{eq:extinction-vs-intensity}) with $n_2 = 2\times10^{-16}$~cm$^2$/W and $\Delta t = 75$~fs. }
    \label{fig:extinction_vs_intensity}
\end{figure}

In order to validate this formula, the extinction factor has been measured as a function of the increasing incident intensity of the pulse $I_{in}$, using the beamsplitter prototype mentioned above with a thickness $e = 6.35$~mm, and with the spatial filter and the spectral filter in the dark output of the interferometer. The measurements are presented in Figure~\ref{fig:extinction_vs_intensity} and show that Equation~\ref{eq:extinction-vs-intensity} is in good agreement with the data.

From this study, we define an upper limit on the intensity of the incident pulse in order to maintain an extinction factor only limited by the beamsplitter asymmetry $\delta a$. For that, we request $\delta \Phi_K < \delta a$, which gives \begin{equation}
    I_{in} \le \frac{\lambda }{2\pi \times n_2 } \times \frac{1}{e}
    \label{eq:intensity_max}
\end{equation}
We note that the upper limit on the intensity is independent of the extinction factor.

Conversely, for a given intensity, the thickness is inversely proportional to that intensity and we obtain an upper limit of the thickness of the beamsplitter:
\begin{equation}
    e \le e_{\mathrm{max}} = \frac{\lambda }{2\pi \times n_2}  \times\frac{1}{I_{in}}
    \label{eq:thickness-max}
\end{equation}

Both conclusions will be used for the optimisation of the parameters in Section~\ref{sec:minimum-waist-at-focus}.


\subsection{Reduction of the interferometer extinction}

\begin{figure}[h!]
    \centering
    \includegraphics[scale=0.4]{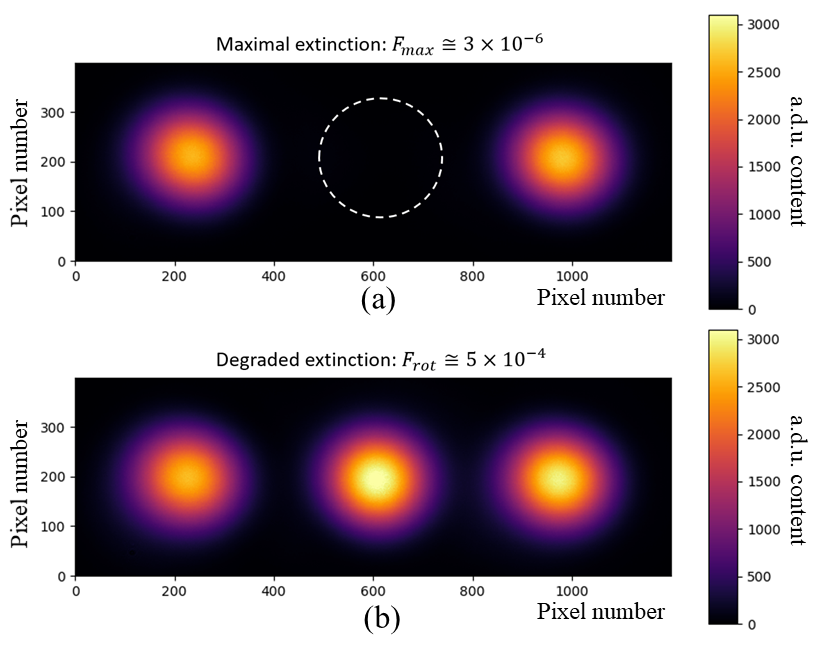}
    \caption{CCD images of the intensity profiles of the interference signal and the back-reflections in the dark output of the interferometer. (a) Intensity profiles at maximal extinction (interference signal in the white dotted area), with $\mathcal{F} _{\mathrm{max}} \simeq 3 \times 10^{-6}$. (b) Intensity profiles with a degraded extinction $\mathcal{F}_{\beta} \simeq 5 \times 10^{-4}$.}
    \label{fig:degradation_extinction}
\end{figure}

With the current available beamsplitter, the presence of the back-reflections means we have to reduce the extinction of the interferometer in order to obtain an interference intensity of the same order as the back-reflection intensities, corresponding to an extinction factor $\mathcal{F} = 5 \times 10^{-4}$.
A typical CCD image recorded in the dark output with a degraded extinction is shown in the lower panel of Figure~\ref{fig:degradation_extinction}

Two different methods were used to degrade the extinction leading to two different configurations of the interferometer with either a low or a high amplification factor $\mathcal{A}$.

\subsubsection*{Low amplification configuration}

In the first configuration, the polarisation of the incident laser pulse is slightly rotated by a small angle $\beta$ with a half-wave plate placed on the probe path, just before entering the interferometer. The electric field $\vec{E}_{in}$ of the incident pulse can be split into two orthogonal $p$ and $s$ polarisation components: $\vec{E}_{in} = \vec{E}_{in,p} + \vec{E}_{in,s}$ with $|\vec{E}_{in,p}| =  |\vec{E}_{in}| \times \cos(\beta)$ and $|\vec{E}_{in,s}| = |\vec{E}_{in}| \times \sin(\beta)$. The degraded extinction becomes dominated by the $s$ component of the beam, due to a large value of the asymmetry coefficient $\delta a_s$ of the beamsplitter for the $s$ polarisation, with an extinction factor $ \mathcal{F}_{\beta} \simeq (\delta a_s)^2 \times \sin^2(\beta)$. 
The $\delta a_s$ coefficient has been calculated by measuring the transmission and reflection coefficients in intensity $T$ and $R$ of the beamsplitter with a pure $s$ polarised incident beam: $T=77.5 \%$ and $R=22.5 \%$, which corresponds to $\delta a_s=0.55$. Rotating the polarisation of the incident beam by an angle $\beta = 40$~mrad leads to the desired reduced extinction factor $\mathcal{F}_{\mathrm{rot}} = 5 \times 10^{-4}$. 
In this configuration, the interference intensity in the dark output is dominated by the $s$ component of the incident beam, and the amplification factor given in Eq.~(\ref{eq:Ion-general}) is strongly reduced, such as: $\mathcal{A}=(1-\delta a_s)/(2\delta a_s)\simeq 1/2$.

\subsubsection*{High amplification configuration}

In the second configuration, the polarisation of the incident beam is now unmodified and maintained purely $p$.
The beamsplitter of the interferometer is rotated by 1 degree in the horizontal plane, changing the incident angle of the laser pulse from 45 to 46 degrees. 
At this incident angle, the measured transmission and reflection coefficients are $R=51 \%$ and $T=49 \%$, corresponding to the asymmetry coefficient $\delta a_p = 0.02$, and an extinction factor $\mathcal{F}_{\mathrm{max}} = (\delta a_p)^2 = 4 \times 10^{-4}$.  
In contrast to the previous configuration, we now have a higher amplification factor $\mathcal{A}=(1-\delta a_p)/(2\delta a_p) \simeq 1/(2\delta a_p) \simeq 24$. 
  
\section{Spatial resolution}
\label{sec:spatial-resolution}

The spatial resolution for the measurement of the barycenter of the intensity profile of the interference signal is limited by three components: the beam pointing fluctuations (which are suppressed by the monitoring of the back-reflections), the phase noise generated by the mechanical vibrations of the interferometer, and finally the inherent shot noise of the CCD camera. These three issues are studied in this section. 

\subsection{The shot noise limit}
\label{sec:shot-noise}

The intrinsic shot noise (or quantum noise) is related to the statistical fluctuations of the average number of photoelectrons $N_{\rm p.e.}$ detected by the CCD camera.
As detailed in~\cite{DeLLight-2021}, the spatial resolution limited by the CCD camera shot noise is independent of the beam width and scales as $d_{\rm pix}/\sqrt{N_c}$, where $d_{\rm pix}$ is the side length of the pixels and $N_c$ is the {\it full well capacity} of the CCD camera, which corresponds to the maximum number of detected photo-electrons per pixel before saturation.

A dedicated test bench has been developed to measure the shot noise of CCD cameras and to measure the ultimate spatial resolution which can be achieved.  
It uses a continuous laser diode beam with a Gaussian transverse intensity profile shaped by a spatial filter. The beam is split into two secondary beams which are sent to the CCD camera. The second beam is used to correct the beam pointing fluctuations and the low frequency drifts. The measurement of the spatial resolution takes into account both the corrections of beam pointing fluctuations and the ON-OFF subtraction procedure defined in section~\ref{sec:beam-pointing-suppression}. The measured value is therefore directly related to the DeLLight spatial resolution $\sigma_y$. 
Details of the test bench and the measurement method are given in~\cite{Max-thesis}.

\begin{figure}[h!]
    \centering
    \includegraphics[scale=0.35]{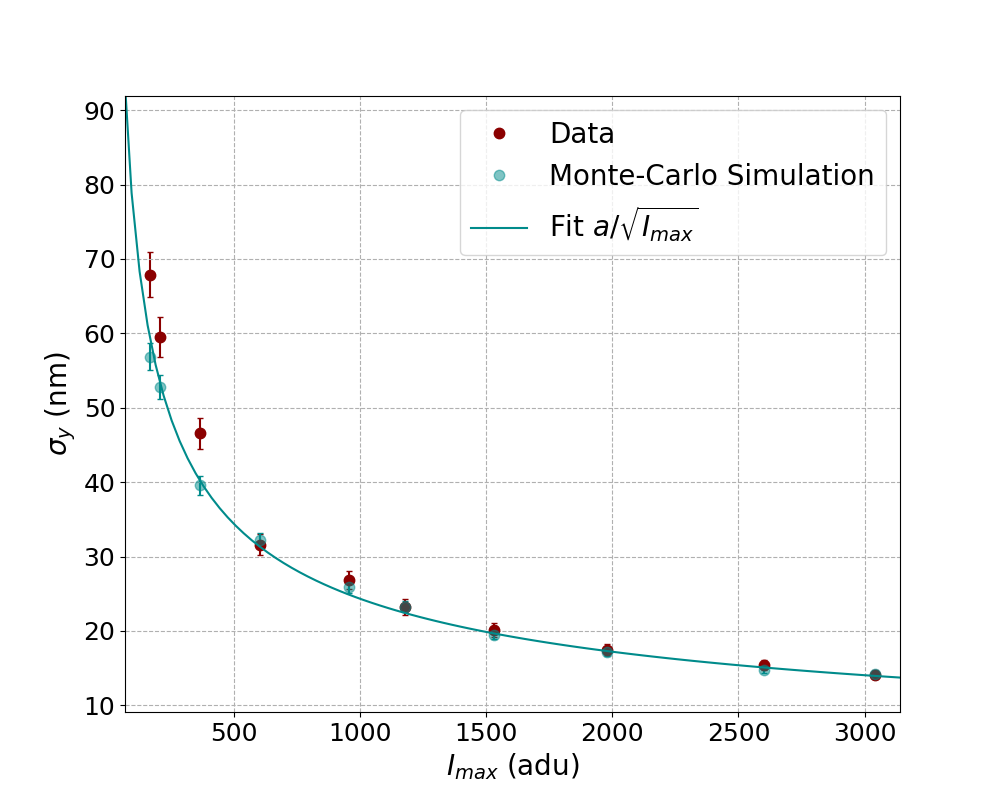}
    \caption{Spatial resolution $\sigma_y$, as a function of the intensity of the pulses $I_{max}$, given in adu content of the pixel at maximum of the intensity profile, measured with the Basler CCD camera model acA4024-29um.
    The red points correspond to the measured values from data. The cyan points correspond to the expected shot noise values calculated by Monte-Carlo simulations. The fitted cyan line corresponds to the expected variation of the shot noise as a function of the square root of the number of detected photo-electrons. }
    \label{fig:shot_noise_vs_I}
\end{figure}

Figure~\ref{fig:shot_noise_vs_I} shows the measured spatial resolution $\sigma_y$ as a function of the beam intensity. Measurements have been performed with the best selected commercial Basler CCD camera which minimizes the ratio $d_{\rm pix}/\sqrt{N_c}$ (model acA4024-29um with pixel size $d_{pixel}=1.8 \mu$m and $N_c=1.1 \times 10^4$ electrons per pixel). 
The measured spatial resolution is in good agreement with the expected shot noise seen in Monte-Carlo simulations.
When the number of photo-electrons recorded in the pixel at maximum of the intensity profile is equal to 3/4 of the full well capacity (typical condition for the DeLLight measurements), we measure a spatial resolution $\sigma_y = 13$~nm.  It is very close to the goal value of 10~nm for the DeLLight project. 
For the Basler CCD used in the current pilot (model acA1920-40gm with pixel size $d_{pixel}=5.86 \ \mu$m and $N_c=3.2 \times 10^4$ electrons per pixel), the measured spatial resolution is $\sigma_y = 30$~nm, again in agreement with the expected Monte-Carlo shot noise.  

\subsubsection*{Requested energy of the incident pulse}
\label{sec:incident-energy}

When limited by the shot noise of the CCD camera, the spatial resolution $\sigma_y$ is then proportional to the transverse beam size $w$ and inversely proportional to the square root of the number of detected photons $N_{\gamma}$ on the camera: $\sigma_y \propto w/\sqrt{N_{\gamma}}$. 
Noting $E_{in}$ the energy of the incident pulse entering the interferometer, $QE$ the quantum efficiency of the CCD camera and $\delta \lambda$ the detected bandwidth of the full bandwidth of the pulse $\Delta \lambda$, we have 
\begin{equation}
N_{\gamma} \propto E_{in} \times \lambda \times \mathcal{F} \times QE \times \frac{\delta \lambda}{\Delta \lambda}
\end{equation}
and the spatial resolution $\sigma_y$ limited by the shot noise is then proportional to:
\begin{equation}
    \sigma_y^2 \propto \frac{w^2}{E_{in}} \times \frac{1}{\mathcal{F} \times QE \times \lambda } \times \frac{\Delta \lambda}{\delta \lambda} 
\end{equation}
For the current pilot interferometer, we have $\sigma_y = 30$~nm with $E_{in} \simeq 2 \ \mu$J, $w=1$~mm, $\mathcal{F} = 5 \times 10^{-4}$, $QE \simeq 0.15$, $\lambda = 800$~nm, $\delta \lambda \simeq 3$~nm and $\Delta \lambda \simeq 30$~nm. 
We obtain a condition on the surface energy density in order to reach the shot noise limit:
\begin{equation}
    \frac{E_{in}}{w^2} = 4.7 \times 10^{-26} \times \frac{1}{\sigma_y^2} \times \frac{1}{\mathcal{F}} \times \frac{1}{QE } \times \frac{1}{\lambda} \times \frac{\Delta \lambda}{\delta \lambda} 
\label{eq:surface_energy_density}
\end{equation}
Thus, with the goal parameters $\mathcal{F} = 10^{-5}$, $QE = 0.8$ and $\delta \lambda = \Delta \lambda$ (no spectral filter) with $\lambda = 800$~nm, we need an incident surface energy  $E_{in}/w^2 \simeq 25 \ \mu$J/mm$^2$ entering the Sagnac interferometer to reach the shot noise spatial resolution $\sigma_y = 10$~nm in the dark output.

\subsection{Suppression of the beam pointing fluctuations}
\label{sec:beam-pointing-suppression}

In the presented measurements, no system of laser beam stabilization is used and significant beam pointing fluctuations are present, leading to fluctuations of the transverse position of the intensity profile in the dark output of the interferometer. 
However, an important feature of the Sagnac interferometer is that the extinction factor in the dark output does not change in the presence of small beam pointing fluctuations. Fluctuations of the beam pointing produce only a simple translation of the intensity profile on the CCD camera, which can be measured and suppressed by monitoring  the position of the interference intensity profile with respect to the back-reflections from the rear side of the beamsplitter.

Here we recall the method, which is similar to the one applied with former prototypes without focusing the beams~\cite{DeLLight-2021}.  
Data of successive odd $(2i+1)$ and even $(2i)$ laser shots are arbitrarily separated into OFF and ON data in order to define an ``ON-OFF'' measurement $i$ (but without pump) using two successive laser shots, at a 5~Hz repetition rate. 
The barycenters of the intensity profiles of the interference signal $\bar{y}_{\mathrm{sig}}$ and the back-reflection $\bar{y}_{\mathrm{ref}}$ are calculated along the vertical 
axis, using a square analysis window (or {\it Region of Interest}, RoI) of size $w_{RoI}$.
As explained in section~\ref{sec:phase-noise}, the presence of the phase noise in the interference signal requires us to reduce the RoI size in order to minimize the phase noise. However, the reduction of the RoI-size tends to decrease the capacity to measure the real displacement of the barycenter. 
We define the efficiency $\epsilon_s(w_{RoI})$ of measuring the $\Delta y$ signal as 
$\epsilon_s(w_{RoI}) = \Delta y_{mes}(w_{RoI})/\Delta y$.
In the following, we use $w_{RoI} = w_{x,y}/2$, where $w_{x,y}$ is the width (FWHM) of the transverse intensity profile, and $\epsilon_s = 0.12$. 

The beam pointing fluctuations are suppressed for each ON and OFF measurement using the correlation of the barycenters of the signal $\bar{y}_{\mathrm{sig}}$ and the back-reflection $\bar{y}_{\mathrm{ref}}$. One obtains the corrected positions:
\begin{eqnarray}
\bar{y}_{\mathrm{corr}}^{\mathrm{OFF}} & = & \bar{y}_{\mathrm{sig}}^{\mathrm{OFF}} - \left( a_{\mathrm{OFF}} \times \bar{y}_{\mathrm{ref}}^{\mathrm{OFF}} + b_{\mathrm{OFF}} \right) \nonumber \\
\bar{y}_{\mathrm{corr}}^{\mathrm{ON}} & = & \bar{y}_{\mathrm{sig}}^{\mathrm{ON}} - \left( a_{\mathrm{OFF}} \times \bar{y}_{\mathrm{ref}}^{\mathrm{ON}} + b_{\mathrm{OFF}} \right) 
\label{eq:BPcorrections}
\end{eqnarray}
where $a_{\mathrm{OFF}}$ and $b_{\mathrm{OFF}}$ are obtained by fitting the linear correlation, using only the OFF measurements. 
The signal $\Delta y (i)$ of the ``ON-OFF'' measurement $i$  is
\begin{eqnarray}
\Delta y (i) = \bar{y}_{\mathrm{corr}}^{\mathrm{ON}}(2i+1) - \bar{y}_{\mathrm{corr}}^{\mathrm{OFF}}(2i).
\end{eqnarray}

\begin{figure*}
    \centering
    \includegraphics[scale=0.33]{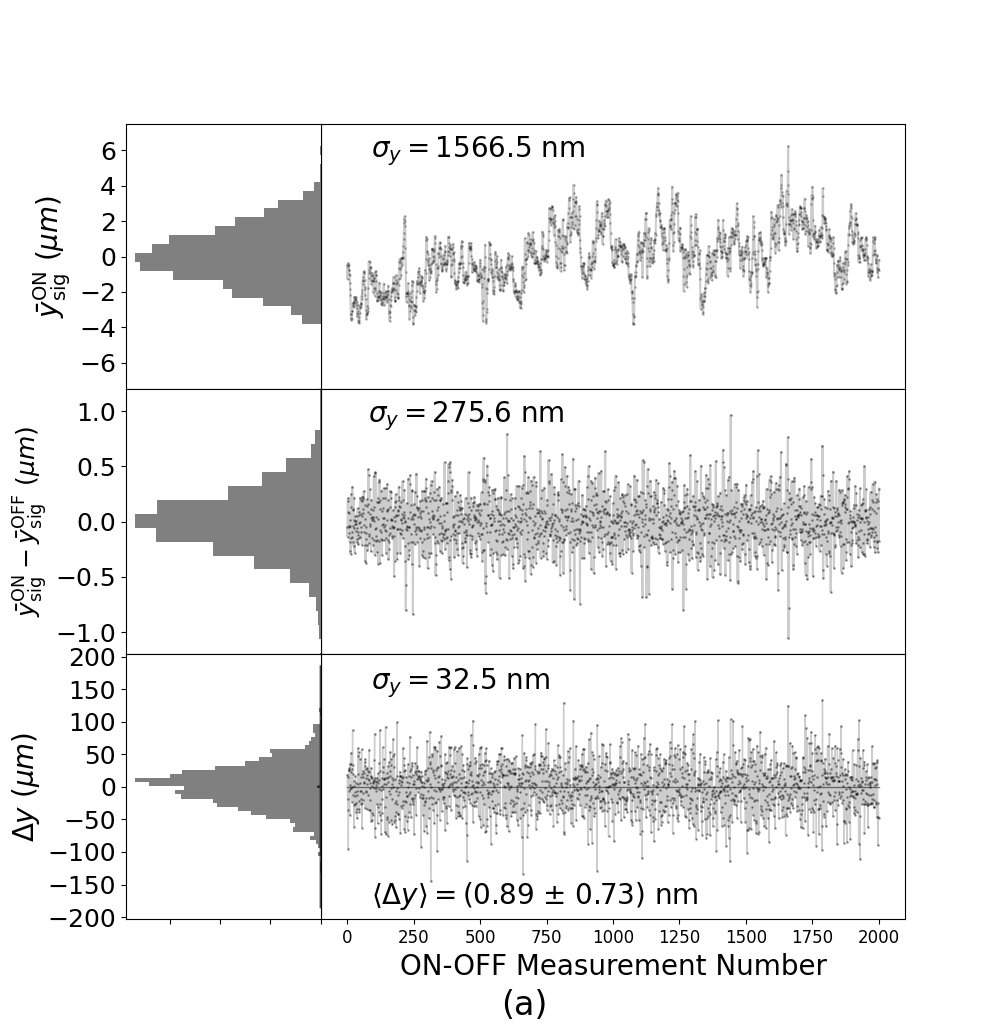}
    \includegraphics[scale=0.33]{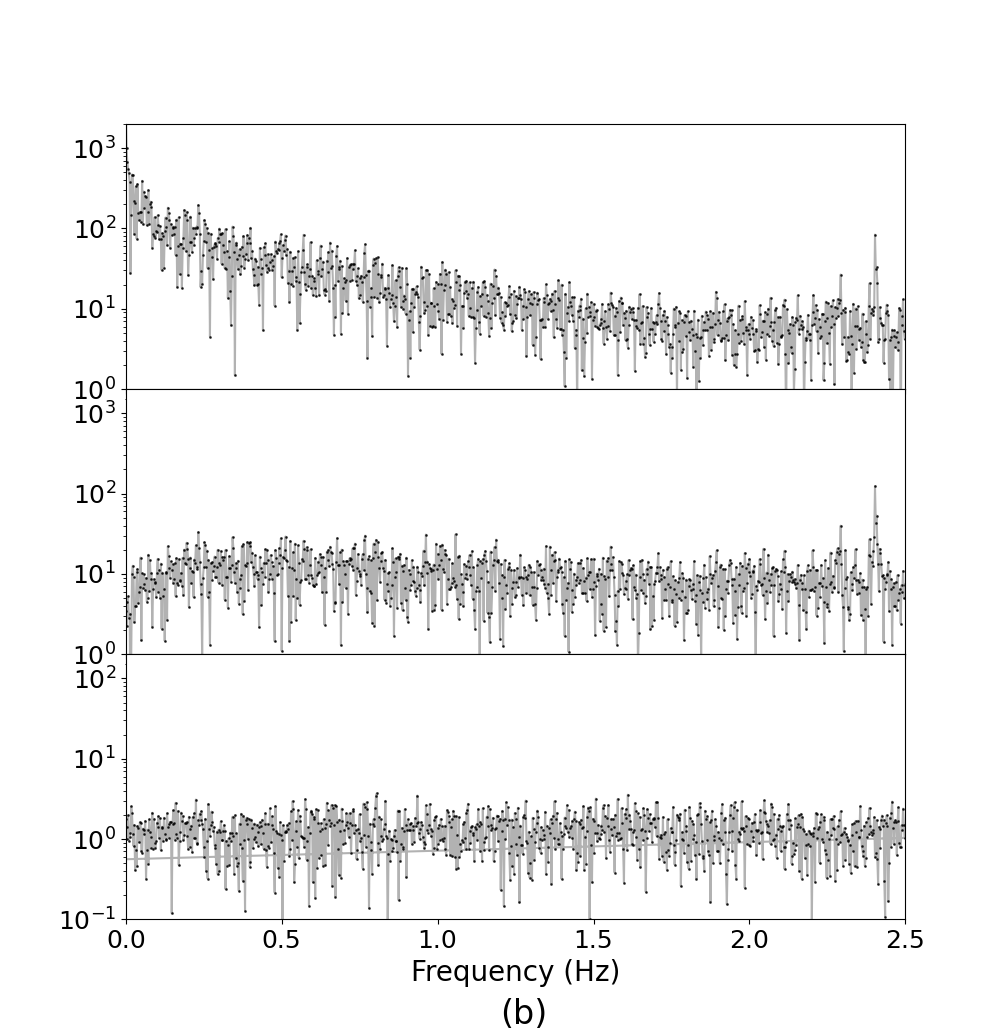}
    \caption{Measurement of the spatial resolution obtained with 4000 successive laser shots at 10 Hz. (a) Distribution of the barycenters in intensity of the interference signal as a function of the ON-OFF measurement $i$ (with $w_{RoI}=w/2$). Upper plot: Raw barycenter position for the OFF data $\bar{y}_{\mathrm{sig}}^{\mathrm{OFF}}(i)$ without any pointing correction. Middle plot: ON-OFF subtraction of the raw barycenter positions $\bar{y}_{\mathrm{sig}}^{\mathrm{ON}}(i) - \bar{y}_{\mathrm{sig}}^{\mathrm{OFF}}(i)$. Lower plot: corrected signal $\Delta y (i)$, after beam pointing correction. The achieved spatial resolution is  $\sigma_y(w_{RoI}=w/2) = 32.5 \pm 0.7$~nm and the average signal is $\Delta y = \langle \Delta y (i) \rangle = 890 \pm 730$~pm, which is compatible with the expected zero value. (b) Corresponding frequency spectra.}
    \label{fig:spatial-resolution-y-freq}
\end{figure*}

We present here the measurement of the spatial resolution, obtained with 4000 successive laser shots (corresponding to almost 7~minutes of measurement at a 10 Hz repetition rate), collected with the pilot interferometer with degraded extinction in the low amplification configuration. 
The distribution of the raw barycenter position of the interference signal for the OFF data, $\bar{y}_{\mathrm{sig}}^{\mathrm{OFF}}(2i)$, is presented in Figure~\ref{fig:spatial-resolution-y-freq} as a function of the ``ON-OFF'' measurement number $i$. 
Beam pointing fluctuations are clearly observed with a poor spatial resolution of about 1.6~$\mu$m (rms). The corresponding frequency spectrum shows a typical $1/f$ drift noise at low frequency and a harmonic peak at  2.4~Hz. 
The ``ON-OFF'' subtraction (at 5~Hz) of the raw barycenter positions $\bar{y}_{\mathrm{sig}}^{\mathrm{ON}}(2i+1) - \bar{y}_{\mathrm{sig}}^{\mathrm{OFF}}(2i)$ acts as a lock-in measurement, suppressing the low-frequency noise. However, the beam pointing fluctuations are still large with a poor spatial resolution of about 300~nm. Finally, the distribution of the signal $\Delta y (i)$, after correction of the beam pointing fluctuations according to Eqs.~(\ref{eq:BPcorrections}), exhibits an excellent spatial resolution $\sigma_y = 32.5 \pm 0.7$~nm, which is in agreement with the CCD shot noise limit as measured with a dedicated test bench and detailed in section~\ref{sec:shot-noise}. The average value over 2000 ON-OFF measurements is $\Delta y = \langle{\Delta y(i)}\rangle = 890 \pm 730$~pm, which is compatible with the expected zero value, with sub-nanometer accuracy. 
The frequency spectrum is flat, as expected for the stochastic shot noise of the CCD camera. 

\subsection{Interferometric noise induced by mechanical vibrations}
\label{sec:phase-noise}

The presence of mechanical vibrations may degrade and limit the spatial resolution.
For a Sagnac interferometer, the interference signal is unaffected by any translation of the beamsplitter or mirrors. 
However, a rotation of the mirrors or the beamsplitter (or a lateral shift of the optical lenses) by an angle $\delta \theta_{\mathrm{noise}}$ along the $y$-axis, generates a relative lateral displacement $\delta y_{\mathrm{noise}}$ in the dark output of the interferometer of the probe and the reference, and by geometric considerations, we have:
\begin{eqnarray}
\delta y_{\mathrm{noise}} = 2 \times L_{opt} \times \delta \theta_{\mathrm{noise}}
\label{eq:shift-noise-vib}
 \end{eqnarray}
where $L_{opt}$ is the effective optical path length between the beamsplitter and the closest optical lens. We have $L_{opt} \approx 35$~cm in the experimental setup\footnote{Since the distance $2 f$ between the two lenses is of the same order as $L_{opt}$, we can assimilate a lateral displacement of the lens into an effective rotation of a mirror.}. 
The rotation of the mirror also produces a rotation of the wavefronts of the two counter-propagating pulses in the dark output, with a relative angle equal to  $2\times \delta \theta_{\mathrm{noise}}$. It produces a phase noise $\delta \Phi_{\mathrm{noise}} (y)$ which is a function of the vertical coordinate $y$:
\begin{eqnarray}
\delta \Phi_{\mathrm{noise}} (y)  = \frac{4\pi}{\lambda_0} \times \delta \theta_{\mathrm{noise}} \times y + \delta \Phi_0
\label{eq:phase-noise-vib}
\end{eqnarray}
where $\lambda_0$ is the central wavelength of the laser.
Finally, using Equations~\ref{eq:Ion-general},~\ref{eq:shift-noise-vib} and \ref{eq:phase-noise-vib}, the intensity profile in the dark output for the  measurement $i$, in the presence of mechanical vibration, is given by:
\begin{align}
\begin{split}\label{eq:I-noise-simu}
I_{\mathrm{out},i}(y) & = (\delta a)^2 \times I_{in}  \left(y+\frac{1-\delta a}{2\delta a} \times \, L_{opt} \times \delta \theta_{\mathrm{noise,i}}\right) \\
 & + \left( \frac{4\pi \, \delta \theta_{\mathrm{noise,i}}}{\lambda_0} \times y + \delta \Phi_0 \right)^2 \times (1-\delta a^2) \\
 & \times I_{in} (y-L_{opt} \, \delta \theta_{\mathrm{noise,i}} )
\end{split}
\end{align}

Since the rotation angle $\delta \theta_{\mathrm{noise}}$ fluctuates between successive laser shots, the spatial barycenter of $I_{\mathrm{OFF}}(y)$ also fluctuates and the spatial resolution is degraded. 
In the low amplification configuration, the amplification factor is $\mathcal{A}=(1-\delta a_s)/(2\delta a_s) \simeq 0.4$ and the barycenter fluctuations are strongly sensitive to the fluctuations of the phase noise $\delta \Phi_{\mathrm{noise}} (y)$.
Conversely, in the high amplification configuration, the amplification factor is $\mathcal{A} = (1-\delta a_p)/(2\delta a_p) \simeq 1/(2\delta a_p) \simeq 25$ and the barycenter fluctuations are dominated by the fluctuations of the amplified displacement $\delta y_{\mathrm{noise}}/(2\delta a_p)$. 

\begin{figure}[h!]
    \centering
    \includegraphics[scale=0.35]{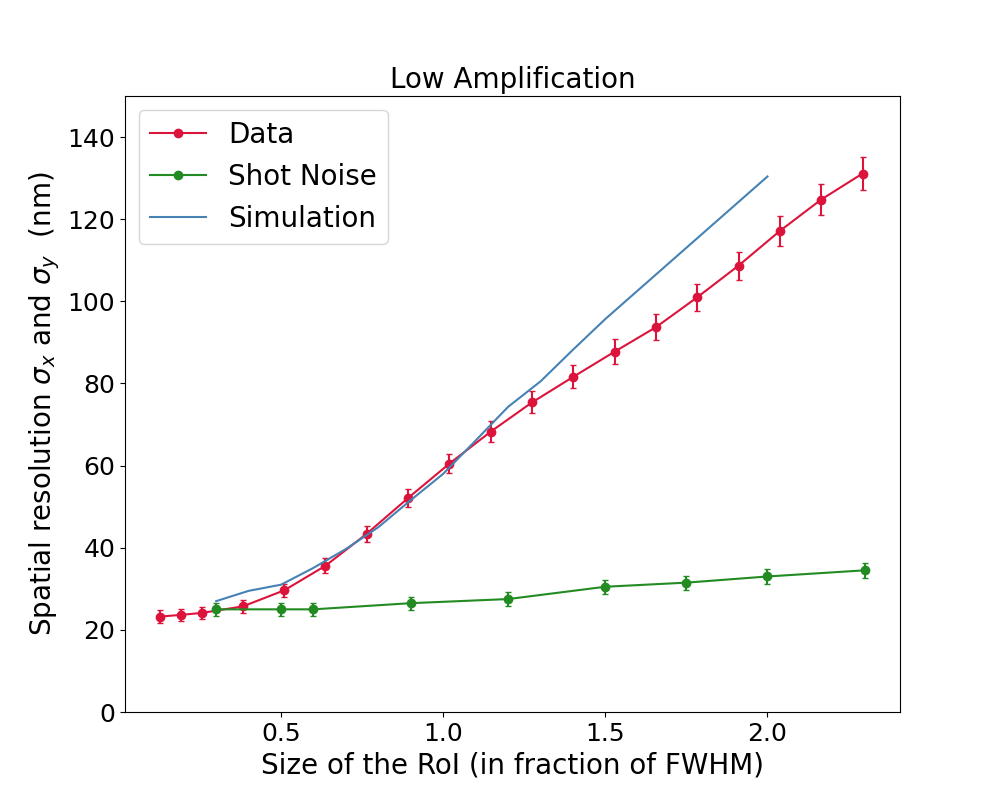}
    \caption{Spatial resolution $\sigma_y$ as a function of the RoI-size $w_{RoI}$ (in FWHM unit) with the low amplification configuration. (Red) Data. (Blue) Monte-Carlo simulation with an angular mechanical vibration $\sigma_\theta = 50$~nrad. (Green) Shot noise of the CCD camera (5.86 $\mu$m) measured on a dedicated test bench.}
    \label{fig:spatial-resolution-vs-roi-low-ampli}
\end{figure}

\begin{figure}[h!]
    \centering
    \includegraphics[scale=0.35]{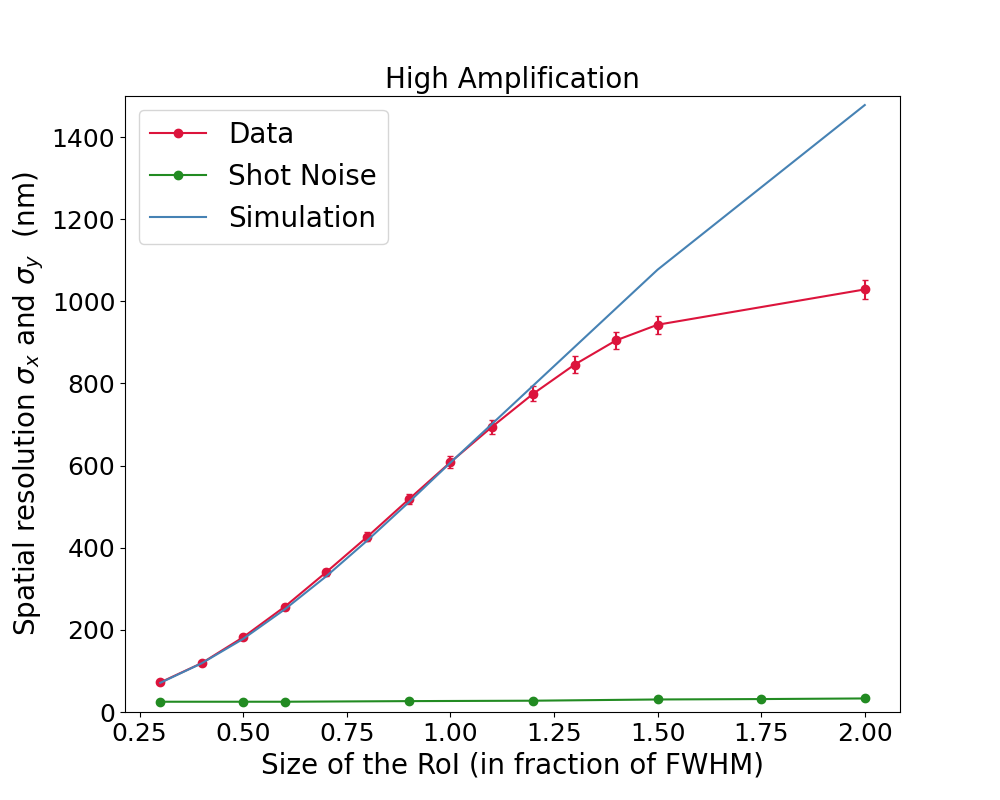}
    \caption{Same as Figure~\ref{fig:spatial-resolution-vs-roi-low-ampli} but in the high amplification configuration and an angular mechanical vibration $\sigma_\theta = 70$~nrad for the Monte-Carlo simulation.}
    \label{fig:spatial-resolution-vs-roi-high-ampli}
\end{figure}

In order to estimate the mechanical vibration $\delta \theta_{\mathrm{noise}}$, we measure the variation of the spatial resolution $\sigma_y$ of the interference intensity profile, as a function of the RoI-size $w_{RoI}$. 
The measurements obtained in the low amplification configuration are presented in Figure~\ref{fig:spatial-resolution-vs-roi-low-ampli}.
In the same figure is also plotted the shot noise resolution measured with the dedicated test bench, as described in section~\ref{sec:shot-noise}. 
The spatial resolution $\sigma_y$ is equal to the shot noise limit for a RoI-size $w_{RoI} \le w_{x,y}/2$, where $w_{x,y}$ is the beam width (FWHM) of the transverse intensity profile. However, for larger RoI-sizes, we observe a degradation of the spatial resolution. 
This degradation is in good agreement with the expected spatial resolution calculated by Monte-Carlo simulation, where we have simulated mechanical vibrations of a mirror with angular gaussian fluctuations $\delta \theta_{\mathrm{noise}} = 50$~nrad (r.m.s.).
The same measurements obtained in the high amplification configuration are presented in Figure~\ref{fig:spatial-resolution-vs-roi-high-ampli}. As expected, the fluctuations are now amplified and the spatial resolution is strongly degraded, even for small RoI-sizes. We verify that the measurements are still in good agreement with the expected spatial resolution calculated by Monte-Carlo simulation, with rotational vibrations on the mirrors with angular Gaussian fluctuations $\delta \theta_{\mathrm{noise}} = 70$~nrad (r.m.s.).
The measured mechanical vibrations are due to the fact that the current interferometer optical board is not isolated from external vibrations.  
A new setup with an improved vibration isolation is under development.
In order to reach the shot noise resolution, while using a large RoI-size (for instance $w_{RoI}=1.5 \times w_{x,y}$, corresponding to an efficiency $\epsilon_s = 70\%$), the requested angular fluctuations must be of the order of $\sigma_\theta \approx 1$~nrad, i.e. a reduction of the vibrations by a factor of approximately 100.

We are also developing a method of high frequency phase noise suppression.
The principle is to split the incident pulse before entering the interferometer into two identical pulses, one being delayed by about 10~ns. We then have two pulses passing through the interferometer, a prompt and a delay pulse. A Pockels cell set in the dark output allows us to record the two pulses spatially separated but simultaneously on the same CCD. At such a high frequency (100 MHz), the phase noise induced by the vibrations is identical for the prompt and the delay pulse. The delay pulse is then used to monitor the phase noise, which is suppressed off-line.  This is similar to the monitoring and suppression of the beam pointing fluctuations using the back-reflections from the rear side of the beamsplitter.

\section{Focusing the probe beam: optimisation of the parameters}
\label{sec:minimum-waist-at-focus}

Ensuring a small  probe waist at focus, $w_0$, is one of the most critical challenges facing the DeLLight experiment. 
For a Gaussian beam, the waist at focus is given by
\begin{equation}
    w_0 = \frac{\lambda_0 \times f}{\pi \times w}
    \label{eq:w0_vs_w}
\end{equation}
where $w$ is the waist of the incident collimated probe beam before focus and $f$ is the focal length.
With the current interferometer, we have $w = 1$~mm and $f=100$~mm, and the minimum waist measured at focus is $w_0 \simeq 25 \ \mu$m. 
The goal value to reach the expected sensitivity of the DeLLight experiment is $w_0 = 5 \ \mu$m. 
A smaller focal length is usually used to reduce the waist at focus, but the DeLLight signal is itself directly proportional to the focal length (see equation~\ref{eq:expected-signal}). Moreover, $f$ also needs to be long enough to guarantee sufficient space for the pump beam to pass through.

Therefore, reaching the requested small waist at focus while keeping a large focal length requires to work with a large incident waist $w$ for the incident collimated beam entering the Sagnac interferometer.
This requires a large beamsplitter thickness $e$, typically $e=2\times w$, in order to separate sufficiently the residual back-reflections from the interference signal in the dark output of the interferometer. 
However, we have seen in section~\ref{sec:limitaion-extinction-factor}, Equation~(\ref{eq:intensity_max}), that the thickness of the beamsplitter is limited by the nonlinear Kerr effect inside the beamsplitter substrate. 
Writing the intensity of the incident pulse $I_{\mathrm{in}}$ as $I_{in}=E_{in}/(2 w^2 \Delta t)$, and by replacing $e=2\times w$, Equation~(\ref{eq:thickness-max}) can be written as an upper limit on the waist of the collimated probe beam:
\begin{equation}
    w < \frac{\lambda \times \Delta t}{2\pi \times n_2}  \times\left(\frac{E_{in}}{w^2}\right)^{-1} 
    \label{eq:emax-1}
\end{equation}
which can be also written as an upper limit on the incident surface energy density entering in the interferometer:
\begin{equation}
    \frac{E_{in}}{w^2} < \frac{\lambda \times \Delta t}{2\pi \times n_2 \times w}
    \label{eq:emax-1bis}
\end{equation}
Moreover, in section~\ref{sec:shot-noise}, Equation~(\ref{eq:surface_energy_density}), we have calculated the requested surface energy density $E_{in}/w^2$ in order to reach the shot noise limit $\sigma_y$. By setting $QE=0.8$ and $\delta \lambda = \Delta \lambda$ (no spectral filter),
Equation~(\ref{eq:surface_energy_density}) becomes
\begin{equation}
    \frac{E_{in}}{w^2}  = 5.9\times10^{-26} \times \frac{1}{\mathcal{F} \times \sigma_y^2 \times \lambda} 
    \label{eq:emax-2}
\end{equation}

\begin{figure}[!h]
    \centering
    \includegraphics[scale=0.35]{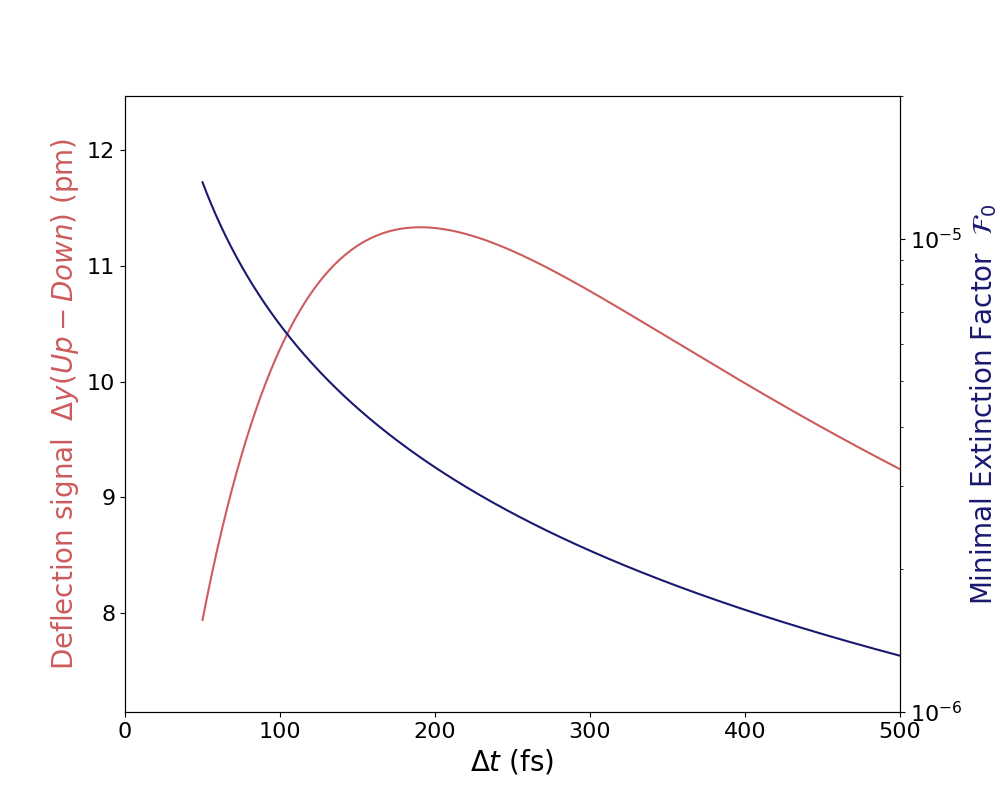}
    \caption{Expected deflection signal $\Delta y(\mathrm{Up-Down})$ (in red) measured in the dark output of the DeLLight interferometer and optimum extinction factor $\mathcal{F}_0$ (in blue) as a function of the pulse duration (fwhm), with $E_{pump}=2.5$~J, $W_0=w_0=5 \ \mu$m, $f=250$~mm ($w = 12.5$~mm), $\sigma_y = 15$~nm and $\theta_{\mathrm{tilt}}=10^{\circ}$.}
    \label{fig:signal_extinction_vs_wmax}
\end{figure}

So we have conflicting constraints:
\begin{itemize}
    \item We need a high extinction (i.e. a low extinction factor $\mathcal{F}$) in order to achieve a high amplification factor.
    \item But we need a high intensity of the interference signal in the dark output to reach the shot noise spatial resolution. This is achieved by entering high incident intensity inside the interferometer, given by Equation~(\ref{eq:emax-2}). 
    \item However, a high incident intensity is going to induce nonlinear Kerr effects inside the beamsplitter, limiting the extinction. This gives a limitation of the incident intensity given by Equation~(\ref{eq:emax-1bis}). 
\end{itemize}
The optimisation of these constraints is obtained by combining Equations~(\ref{eq:emax-1bis}) and (\ref{eq:emax-2}). It gives the optimum value of the  extinction factor $\mathcal{F}_0$, needed to reach the shot noise spatial resolution while avoiding saturation of the extinction due to the Kerr effect in the beamsplitter: 
\begin{equation}
    \mathcal{F}_0 = 3.7 \times 10^{-25} \times \frac{n_2}{\sigma_y^2 \times \lambda^2 \times \Delta t } \times w
    \label{eq:minimal-extinction-factor}
\end{equation}
Combining this equation with Equations~(\ref{eq:expected-signal}) and (\ref{eq:w0_vs_w}), we finally obtain the expected signal $\Delta y$ as a function of the incident beam size $w_{\mathrm{max}}$:
\begin{align}
\begin{split}\label{eq:Deltay_vs_wmax}
    \Delta y_{\mathrm{QED}} = & \ 3.1 \times 10^{11} \times E_{pump} \times \frac{w_0}{(w_0^2+W_0^2)^{3/2}}  \\
    & \times \sigma_y \times r_{\mathrm{tilt}}(\Delta t) \times \sqrt{\Delta t} \times \sqrt{w}
\end{split}
\end{align}
Figure~\ref{fig:signal_extinction_vs_wmax} shows the expected signal $\Delta y(\mathrm{Up-Down}) = 2\times \Delta y_{\mathrm{QED}}$, and the corresponding optimum extinction factor $\mathcal{F}_0$ as a function of the pulse duration $\Delta t$, for a minimum waist at focus $W_0 = w_0 = 5 \ \mu$m for the pump and the probe beams, for a focal length $f=250$~mm (corresponding to an incident beam size $w=12.5$~mm), a tilt angle $\theta_{\mathrm{tilt}}=10^{\circ}$, and a spatial resolution $\sigma_y = 15$~nm (corresponding to the shot noise measured with our most appropriate CCD camera). The signal reaches a maximum value $\Delta y(\mathrm{Up-Down}) \simeq 12$~pm when the pulse duration is $\Delta t \simeq 180$~fs. The associated optimum extinction factor is $\mathcal{F_0} = 4 \times 10^{-6}$. 

Another option is to reduce by a factor of 2 the wavelength of the incident laser pulse using second harmonic generation (from $\lambda = 800$~nm
to $\lambda = 400$~nm) before entering the interferometer. 
The expected signal given in Equation~(\ref{eq:Deltay_vs_wmax}) is independent of the wavelength. However, from Equation~(\ref{eq:minimal-extinction-factor}), it allows to work with an extinction factor 4 times larger. Another advantage is to reduce by a factor of 2 the diameter $w$ of the probe beam, while maintaining the small
waist at focus. 

It is also worth noting from Equation~(\ref{eq:Deltay_vs_wmax}) that the sensitivity of the signal measurement, given by the ratio $\Delta y/\sigma_y$, does not depend on the spatial resolution $\sigma_y$. However, from Equation~(\ref{eq:minimal-extinction-factor}), working with a degraded spatial resolution (a larger $\sigma_y$ value) would require the use of a stronger extinction (a lower extinction factor $\mathcal{F}$) in order to maintain the same sensitivity.



Finally, we compare in Table~\ref{tab:sensitivity} the expected sensitivity of the DeLLight experiment using the LASERIX facility, with two other femtosecond laser facilities under development: the KALDERA laser (DESY) with an energy of 3~J per pulse with 1~kHz repetition rate~\cite{{KALDERA}}, and the HAPLS laser (ELI Beamline center) with an expected energy of 30~J per pulse with 10~Hz repetition rate~\cite{HAPLS}. While the expected sensitivity (1~$\sigma$ confidence level) to observe the QED signal could be reached after about 4~days of collected data with the LASERIX facility, it could be reached in only half an hour with the KALDERA or the HAPLS lasers.

\begin{table}[!h]
\centering
\begin{tabular}{c|c|c|c|c}
\hline
\hline
Laser       & Energy   	&   Repetition  &  $\Delta y$  &  $T_{\mathrm{obs}}$ 	  \\
			&    		&   Rate  		&   	(Up-Down)       &  (1$\sigma$ sensitivity) \\
\hline
\hline
LASERIX     &  2.5 J    &  10 Hz        &  12 pm      &  4 days              	\\
KALDERA     &  3 J      &  1 kHz        &  15 pm        &  36 min.              	\\
HAPLS       &  30 J     &  10 Hz        &  0.15 nm     &  36 min.              	\\
\hline
\hline
\end{tabular}
\caption{Expected QED DeLLight signal $\Delta y(\mathrm{Up-Down})$ and required duration of measurement $T_{\mathrm{obs}}$ in order to reach 1~$\sigma$ sensitivity for various laser facilities.}
\label{tab:sensitivity}
\end{table}

\section{Conclusion}
\label{sec:conclusion}

A Sagnac interferometer working with femtosecond laser pulses 
has been developed for the DeLLight project, whose aim is to measure by interferometry the deflection in vacuum of a low energy probe pulse crossing the vacuum index gradient induced by the interaction with a high energy pump pulse. 
This new pilot interferometer includes the focus of the circulating probe pulses needed for the pump-probe interaction.
In the present study,  we have measured and characterized the critical experimental parameters limiting the sensitivity of the DeLLight experiment, namely the extinction factor $\mathcal{F}$, the spatial resolution $\sigma_y$  and the minimum waist at focus $w_0$ of the probe beam inside the interferometer. 

An extinction factor $\mathcal{F}=3 \times 10^{-6}$ has been successfully obtained thanks to the addition of a spatial filter in the dark output of the interferometer. It corresponds to the optimal extinction to reach the expected sensitivity. However, it has been obtained for a restricted spectral bandwidth and it is still limited by the back reflections of the beamsplitter. The development of a new dedicated beamsplitter is in progress to solve these limitations.
We have also measured that the extinction is inherently limited by the optical nonlinearities inside the beamsplitter. 

We have shown that the spatial resolution in the dark output of the interferometer is currently limited by the phase noise fluctuations induced by mechanical vibrations of the interferometer. 
The phase noise must be reduced by two orders of magnitude in order to reach the ultimate shot noise resolution. To achieve this, a method of high frequency (100~MHz) phase noise suppression is being developed. It consists of splitting the incident pulse, before entering the interferometer, into two identical prompt and delay pulses. The delay is used to monitor and suppress the phase noise in a similar way to the monitoring and suppression of the beam pointing fluctuations already used in the DeLLight interferometer. 

With the current interferometer, the minimum waist measured at focus is $w_0 \simeq 25 \ \mu$m. It is 5 times larger than the requested value $w_0 = 5 \ \mu$m to reach the expected sensitivity of the DeLLight project. This can be achieved by increasing the transverse size of the incident beam entering the Sagnac. However, we have shown that two conditions constrain the experimental parameters: the extinction limited by nonlinearities inside the beamsplitter and the required number of detected photons on the CCD to reach the shot noise spatial resolution. After optimisation of the parameters we have shown that the expected sensitivity (1~$\sigma$ confidence level) to observe the QED signal could be reached after about 4 days of collected data with the LASERIX facility, while it could be reached in only half an hour with the KALDERA or the HAPLS lasers.

We finally mention that a proof of concept of the DeLLight experiment has been recently achieved by measuring, with the current DeLLight interferometer, the deflection of a probe pulse induced by a low-energy pump pulse via the Kerr effect in air. Results of these measurements, presented in a separate article~\cite{DeLLight-in-air}, validate the DeLLight interferometric amplification method.

\section*{Acknoledgment}
This work is funded by the French National Research Agency via Grant No. ANR-22-CE31-0003-01.
 



\end{document}